# Compatibilities and supercompatibility conditions in shape memory alloys determined from correspondence, metrics and symmetries

Cyril Cayron, EPFL, STI, LMTM, Rue de la Maladière 71b, 2000 Neuchâtel, Switzerland

cyril.cayron@epfl.ch

## Abstract

The phenomenological theory of martensite crystallography (PTMC) developed in the 1950's explains the main crystallographic and microstructural features of martensite in shape memory alloys, such as the habit planes of bi-variant laminate martensite product, and the transformation twins between the variants. It also permits to determine the austenite and martensite lattice parameters that allow supercompatibility, which has driven important research and development of new shape memory alloys with low hysteresis and high cyclability. Supercompatibility takes the form of three mathematical equations called "cofactor conditions". The calculations are in great part based mathematical tools from continuum mechanics (polar decompositions and stretch tensors). They were recently replaced by pure crystallographic tools (metric tensors, group of symmetries and correspondence) in an alternative approach called correspondence theory (CT). The CT allows for the direct calculation of the transformation twins and their generic and non-generic characteristics. These twins ensure the compatibility at martensite/martensite (M/M) junction planes. Here, we show that the CT can also be used to determine the conditions of austenite/martensite (A/M) compatibility, and A/M/M supercompatibility. After a brief reminder of the CT, a new symmetric matrix called "compatibility by metrics correspondence" (CMC) is introduced. The A/M compatibility condition is obtained when the double cone based on the quadratic form of the CMC degenerates into a double or simple plane (the habit plane). With such a condition, the $A \rightarrow M$ lattice distortion is an invariant plane strain (IPS), and its shear direction can be deduced from the shear plane by using a second symmetric matrix called SMC for "shear by metric correspondence". The A/M/M supercompatibility is obtained when the shear direction of the IPS and the shear direction of the twin become proportional and linked by a shear/shear equation. The CT approach was applied to B19' martensite in NiTi alloys. It was shown that this monoclinic phase is supercompatible for lattice parameters that are analytical function of the monoclinic angle. A supercompatible B19' martensite was found with lattice parameters close to those of the actual B19' martensite observed in binary NiTi alloys. It was checked numerically that all the supercompatible solutions for B19' determined by CT also verify the cofactor conditions. However, a full and formal equivalence between the CT and the PTMC could not yet be established. Since the CT is based on pure crystallography tools and direct geometrical considerations, it may represent a good alternative to the PTMC for future researches that would aim at targeting supercompatible martensite for new shape memory alloys with high cyclability.

## 1 Introduction

The phenomenological theory of martensite crystallography (PTMC) is a cornerstone of physical metallurgy for understanding the martensitic microstructures. It dates from 1950's and was initially aimed at explaining the orientation relationships and the habit plates observed in martensitic steels [1,2]. Since then, the theory has been successfully extended to cover the martensite in other alloys, such as the shape memory alloys [3,4]. The PTMC is built on three fundamental hypotheses: (1) A correspondence should exist between the austenite and martensite lattices. The correspondence matrix specifies in which crystallographic direction of the martensite phase a direction of the parent austenite phase is transformed. This correspondence requires a crystallographic model of the transformation, such as the Bain model proposed in 1924 for the face centred cubic (fcc) to body centred cubic (bcc) martensitic transformation in steels [5]. It is important to note that it is actually the stretch matrix $\mathbf{U}$ deduced from the correspondence that is used for the PTMC calculations, and not the correspondence matrix itself. (2) The habit plane (HP), which is the interface plane between the martensite product (lath, plate, lenticle), should remain unrotated and undistorted by the macroscopic deformation. This implies that the shape macroscopic shape strain $\mathbf{P}$ associated with the formation of the martensite product must be an invariant plane strain (IPS). A third hypothesis is required; the mathematical form of the shape strain depends on the version of PTMC: (3a) In the version developed by Bowles and Mackenzie (BM) [1], see also [6], a complementary simple shear matrix $\mathbf{S}$ called "lattice-invariant shear" (LIS) is introduced. Its is associated with the stretch matrix $\mathbf{U}$ and a free rotation $\mathbf{Q}$ in order to obtain the IPS by the equation $\mathbf{P} = \mathbf{S}^{-1}\,\mathbf{Q}\,\mathbf{U}$. The name LIS comes from the fact that this complementary shear was initially assumed to be dislocation slip, but deformation twinning can also be proposed, and for such cases, the lattice is not invariant and the name LIS may be misleading. (3b) In the version developed by Weschler, Lieberman and Read (WLR) [2], the IPS macroscopic deformation is obtained by a combination of two twin-related variants. The two PTMC versions are equivalent when the LIS of the BM version is chosen to be the shear associated with the twin of the WLR version [7]; however, contrarily to the BM version, the WLR version does not need any assumption on the LIS because the twins are outputs. These twins are not "deformation twins" that can be arbitrarily introduced; they are called "transformation twins" whose characteristics (shear plane and direction) are obtained by calculations. The WLR version has been mathematically developed for decades [4,8,9] and applied with success to many shape memory alloys [10,11].

The method and the main equations of the WRL version are now briefly recalled. The first step is to establish a crystallographic model in order to determine the correspondence matrix. Let us take as example, the Bain correspondence for the fcc austenite (A) to bcc martensite (M) transformations in steels [5] that can written $\frac{1}{2}[110]_A \rightarrow [100]_M$, $\frac{1}{2}[1\bar{1}0]_A \rightarrow [010]_M$ and $[001]_A \rightarrow [001]_M$. Note that the arrows mean that the directions that are in correspondence are not necessarily parallel. From the correspondence, a stretch matrix can be calculated. Here, $\mathbf{U}$ is not only symmetric; it is diagonal, $\mathbf{U} = \mathrm{diag}\left(\frac{a_M}{\frac{a_A}{\sqrt{2}}}\ \ \frac{a_M}{\frac{a_A}{\sqrt{2}}}\ \ \frac{a_M}{a_A}\right)$. In steels, the first and second terms in the diagonal are larger than 1 (extension) and the third one is smaller than 1 (contraction). The stretch matrix is a key component of the lattice distortion $\mathbf{F}$ between austenite and martensite because $\mathbf{F}$ can always be expressed as a combination of a rotation $\mathbf{Q}$ and a symmetric (stretch) matrix $\mathbf{U}$ by polar decomposition $\mathbf{F} = \mathbf{Q}\,\mathbf{U}$. Polar decomposition implies to create an orthonormal basis in which the three vectors of conventional crystallographic bases of the phases are decomposed. Once the stretch matrix $\mathbf{U}$ is determined, its variants $\mathbf{U}_i$ are calculated by considering the symmetries of the parent phase. We insist on the fact $\mathbf{U}_i$ are "stretch variants"; it is not correct to call them "correspondence variants". The distortion matrices $\mathbf{F}_i = \mathbf{Q}_i\,\mathbf{U}_i$ and $\mathbf{F}_j = \mathbf{Q}_j\,\mathbf{U}_j$ of two variants $i$ and $j$, respectively, are "compatible" when they are twin-related. This relation was mathematically formalized by Ball and James [8] and Bhattacharya [4] as a rank-1 condition: there is a plane of normal $\mathbf{n}$ and a direction $\mathbf{a}$ in this plane such that $\mathbf{F}_i - \mathbf{F}_j = \mathbf{a} \otimes \mathbf{n}$. The twin plane $\mathbf{n}$ and the twin direction $\mathbf{a}$ are called "shear plane" and "shear direction", respectively, as if the transformation twin could be identified to a deformation twin, which is debatable. The rank-1 equation is solved by calculating the eigenvalues $\chi_1$, $\chi_2$, $\chi_3$ and eigenvectors $\mathbf{e}_1$, $\mathbf{e}_2$, $\mathbf{e}_3$ of the matrix $\mathbf{F}_j{}^{-\mathrm{t}}\,\mathbf{F}_i{}^{\mathrm{t}}\,\mathbf{F}_i\,\mathbf{F}_j{}^{-1} = \mathbf{U}_j{}^{-\mathrm{t}}\,\mathbf{U}_i{}^{\mathrm{t}}\,\mathbf{U}_i\,\mathbf{U}_j{}^{-1}$, with "t" for "transpose". Some solutions exist for $\mathbf{a}'$ and $\mathbf{n}$ if and only if $\chi_1 \leq 1$, $\chi_2 = 1$, $\chi_3 \geq 1$. The habit plane of a martensite product constituted of the two variants $i$ and $j$ is then calculated by assuming that the macroscopic deformation is an IPS, written $\bar{\mathbf{F}} = f\,\mathbf{F}_i + (1-f)\,\mathbf{F}_j = \mathbf{I} + \mathbf{b} \otimes \mathbf{m}$, where $f$ represents the volume faction of each variant between 0 and 1, and $\mathbf{I}$ is the identity matrix. The habit plane $\mathbf{m}$ is the invariant plane of the IPS. The calculations of $f$, $\mathbf{b}$, $\mathbf{m}$ are quite long (the details are skipped here).

There are specific cases of martensitic transformations in which pairing the variants is not necessary to obtain a macroscopic IPS deformation. Indeed, individual variants can have a coherent interface with austenite if **F** is already an IPS, and no transformation twins are required. This implies very specific relations between the lattice parameters of the austenite and martensite phases. In its modern mathematical form [4,8], the IPS condition is generally formulated by the equation $\lambda_2 = 1$, where $\lambda_1 \leq \lambda_2 \leq \lambda_3$ are the eigenvalues of the stretch matrix $\mathbf{U}$. Twenty years ago, Cui *et al.* [12] investigated the resistance to the thermal fatigue of ternary NiTi alloys and could show that the thermal hysteresis is significantly reduced when $\lambda_2$ becomes close to 1. This study also permitted to give up the idea that the volume change could affect the reversibility, since no correlation were found between the thermal hysteresis and $\det(\mathbf{U}) = \lambda_1 \lambda_2 \lambda_3$. A few years later, Delville *et al.* [13] showed in NiTiPd alloys that the martensite morphology evolves from twinned lamellae to twinless plate as $\lambda_2$ approaches 1. The condition $\lambda_2 = 1$ is necessary but not totally sufficient to reach "supercompatibility". A phase transformation is said to respond to the supercompatibility conditions when martensite can be formed in any volume fraction $f$ of two variants *i* and *j*. These conditions were extracted from the mathematical development made by Ball and James [8] and initially stated by James and Zhang [14]: " *we call the cofactor conditions, at which an even more spectacular "accident" of compatibility occurs. The cofactor conditions presuppose that* $\lambda_2 = 1$, *and they also depend on the choice of the twin system,* $\boldsymbol{a}, \boldsymbol{n}$". The vectors $\mathbf{a}$ and $\mathbf{n}$ are the "shear" direction and the normal to "shear" plane. The supercompatibility conditions are thus the association of the IPS condition $\lambda_2 = 1$, and two additional conditions, an equality and an inequality, implying the twin elements ($\mathbf{a}$ and $\mathbf{n}$). The three supercompatibility conditions (SC) are:

$$\text{SC1} : \lambda_2 = 1 \tag{1}$$

$$\text{SC2} : \mathbf{a}.\mathbf{U}\,\mathrm{cof}(\mathbf{U}^2 - \mathbf{I})\,\mathbf{n} = 0 \tag{2}$$

$$\text{SC3} : \mathrm{tr}(\mathbf{U}^2) - \det(\mathbf{U}^2) - \frac{\mathbf{a}^2}{4} - 2 \geq 0 \tag{3}$$

The second one, SC2, is called cofactor condition (CC). The notation "CCI" and "CCII" is often used to distinguish whether SC2 is obtained from a type I or a type II twin [15]. Zhang *et al.* [16] showed that the cofactor condition is equivalent to another condition that states that the unit vector $\hat{\mathbf{e}}$ parallel to the twinning plane normal (for type I twins) or to the 180° rotation axis (for type II twins) should verify the condition $||\mathbf{U}^{-1}\,\hat{\mathbf{e}}|| = 1$, for type I twins, or $||\mathbf{U}\,\hat{\mathbf{e}}|| = 1$, for type II twins. Recent reviews of the supercompatibility conditions and their effects on the thermal hysteresis can be found in Refs. [15,17,18]. All the mentioned studies have opened a new domain in the science of shape memory alloys called "phase engineering" [15] in which additional elements are added to the alloys in order to reach the supercompatibility conditions. There is still however a lack of understanding on the physical meaning of these conditions. The conditions SC2 (or $X_I$ and $X_{II}$) and SC3 result from calculations that are not always easy to follow or to understand geometrically. Gu *et al.* [15] admitted to "*not understand the relative roles of* $\lambda_2 = 1$ *vs. the full cofactor conditions in determining hysteresis and reversibility*". Actually, even the meaning of $\lambda_1, \lambda_2, \lambda_3$ the eigenvalues of $\mathbf{U}$ can be difficult to grasp, as the same authors noticed that the positions of the points $(\lambda_1, \lambda_3)$ for cubic to orthorhombic transformation for different alloys "*fall closely on a straight line in this plot, a fact that is not understood"* [15] .

The present work aims at showing that the PTMC calculations based on polar decompositions and stretch matrices can be substituted by more direct and comprehensive calculations based on the metrics and group of symmetries of the parent and daughter phases, and on the correspondence between the two phases. First, we will show by simple geometry that the supercompatibility conditions SC1, 2, 3 can be substituted by other conditions that are easier to understand: austenite/martensite (A/M) and martensite/martensite (M/M) compatibilities, and an "shear/shear" equation that links the shear direction of the martensitic lattice distortion to the shear direction of the transformation twin. Incidentally, we will explain why the $(\lambda_1, \lambda_3)$ experimental points seem to fall closely on a straight line. Second, we will give a brief summary of the Correspondence Theory (CT) introduced a few years ago [19]. We will recall how the transformation twins can be calculated directly from the parent and daughter metrics and from the correspondence matrix, and how the symmetries should be taken into account to avoid redundant calculations. Third, a new method to determine the lattice parameters that verify supercompatibility will be proposed. The calculations to determine the A/M compatibility will be based on a matrix called "compatibility by metric correspondence". We will show that a coherent A/M interface exists if its quadratic form matrix is degenerated into a double or simple plane. Different orders of degeneracy will be distinguished. The lattice distortion shear direction can then be deduced from the shear plane of the IPS by using a second matrix called "shear by metric correspondence". Supercompatibility is obtained when the shear/shear equation is verified. An example of supercompatibility conditions obtained by the CT will be given in the case of B2 $(cubic) \rightarrow$ B19$'(monoclinic)$ martensite transformation in NiTi alloys. The similarities and differences between the CT and PTMC theories will be eventually discussed.

# 2 Geometrical interpretation of the supercompatibility conditions

## 2.1 The link between the stretch and the shear values

The condition SC1, $\lambda_2 = 1$, implies that the lattice distortion can form a coherent interface with the parent austenite and therefore be an IPS. Let us show it. As discussed in the introduction, any lattice distortion contains a symmetric stretch component $\mathbf{U}$ written in a reference orthonormal basis. We note $(\mathbf{e}_1, \mathbf{e}_2, \mathbf{e}_3)$ the eigenvectors of $\mathbf{U}$. They form an orthonormal basis in which the stretch is simply the matrix $\mathbf{U} = \mathrm{diag}(\lambda_1, \lambda_2, \lambda_3)$. In particular, since $\lambda_2 = 1$, $\mathbf{U}\,\mathbf{e}_2 = \mathbf{e}_2$. Since $\lambda_1 \leq \lambda_2 \leq \lambda_3$ , it exists a vector $\mathbf{v}$ in the plan $(\mathbf{e}_1, \mathbf{e}_3)$ such that $\|\mathbf{U}\,\mathbf{v}\| = \|\mathbf{v}\|$. We note $\mathbf{v}' = \mathbf{U}\,\mathbf{v}$, and $\mathbf{R}_v$ the rotation of angle $-(\widehat{\mathbf{v}',\mathbf{v}})$ around $\mathbf{e}_2$ that compensates the rotation of $\mathbf{v}$. Since $\mathbf{R}_v\,\mathbf{U}\,\mathbf{e}_2 = \mathbf{e}_2$ and $\mathbf{R}_v\,\mathbf{U}\,\mathbf{v} = \mathbf{v}$, the lattice distortion $\mathbf{F} = \mathbf{R}_v\,\mathbf{U}$ is an IPS, and its shear plane is $(\mathbf{e}_2, \mathbf{v})$. It is the plane of perfect austenite/martensite coherency.

The generic form of an IPS restricted to the 2D space normal to $\mathbf{e}_2$ is a matrix made of the pure shear amplitude $\tau$ and dilatation $\delta$ normal to the shear plane given by

$$\mathbf{F} = \begin{pmatrix} 1 & \tau \\ 0 & 1+\delta \end{pmatrix} = \mathbf{I} + \mathbf{d}\,\mathbf{m}^{\mathrm{t}} \tag{4}$$

The dyadic product $\mathbf{d}\,\mathbf{m}^{\mathrm{t}}$ notation is preferred to its equivalent product $\mathbf{d} \otimes \mathbf{m}$ because it is directly compatible with conventional matrix product rules used in the paper. Note that all the vectors in the equations are column vectors by default whatever their space (direct or reciprocal), even if, for safe of space, they are often written in line in the text. The sign "transpose" is used when the vector is written in line in the equation. Equation (4) is the expression of an IPS on a horizontal plane $\mathbf{m} = (0,1)$ along a shear direction $\mathbf{d} = \begin{bmatrix} \tau \\ \delta \end{bmatrix}$. The inverse of $\mathbf{F}$ is also an IPS; it is given by

$$\mathbf{F}^{-1} = \begin{pmatrix} 1 & -\dfrac{\tau}{1+\delta} \\ 0 & \dfrac{1}{1+\delta} \end{pmatrix} = \mathbf{I} - \frac{1}{1+\delta}\mathbf{d}\,\mathbf{m}^{\mathrm{t}} \tag{5}$$

The eigenvalues of $\mathbf{F}^{\mathrm{t}}\,\mathbf{F} = \mathbf{U}^2$ noted $\mu_i$ are the square of the eigenvalues of $\mathbf{U}$, i.e. $\mu_i = \lambda_i^2$, and solutions of the quadratic form $\mu^2 - (1 + (1+\delta)^2 + \tau^2)\,\mu + (1+\delta)^2 = 0$. Thus, the values $\lambda_i$ are

$$\lambda_1 = \sqrt{\mu_1} = \frac{\sqrt{1+(1+\delta)^2+\tau^2-\sqrt{\Delta}}}{\sqrt{2}} \text{ and } \lambda_3 = \sqrt{\mu_3} = \frac{\sqrt{1+(1+\delta)^2+\tau^2+\sqrt{\Delta}}}{\sqrt{2}} \text{ with } \Delta = (\delta^2+\tau^2)((2+\delta)^2+\tau^2)$$

The easiest way to change the parameters $(\lambda_1, \lambda_3) \leftrightarrow (\tau, \delta)$ is to use the sum $\mu_1 + \mu_3$ and product $\mu_1\mu_3$ by

$$\lambda_1^2 + \lambda_2^2 + \lambda_3^2 = \lambda_1^2 + \lambda_3^2 + 1 = \mathrm{tr}\,(\mathbf{U}^2) = 2 + (1+\delta)^2 + \tau^2 \tag{6}$$

$$\lambda_1^2\,\lambda_2^2\,\lambda_3^2 = \lambda_1^2\,\lambda_3^2 = \det(\mathbf{U}^2) = (1+\delta)^2 \tag{7}$$

The volume change is $\frac{V'}{V} = \det(\mathbf{F}) = \det(\mathbf{U}) = 1 + \delta$. To the author's knowledge, in all the martensitic phase transformations reported in literature, the dilatation part $\delta$ is significantly smaller than the unit. Consequently, equality (7) can be approximated by $\lambda_3 = \frac{1+\delta}{\lambda_1} \approx \frac{1}{\lambda_1}$. This approximate inverse relation between the two eigenvalues appears clearly in Fig.3 of Gu et al. [15], as shown in Figure 1. The fact that $\delta \ll 1$ and the Taylor expansion $\frac{1}{1-(1-\lambda_1)} \approx 1 + (1 - \lambda_1)$ for $\lambda_1 \approx 1$ explains that the experimental points $(\lambda_1, \lambda_3)$ "*fall closely on a straight line in this plot*" [15].

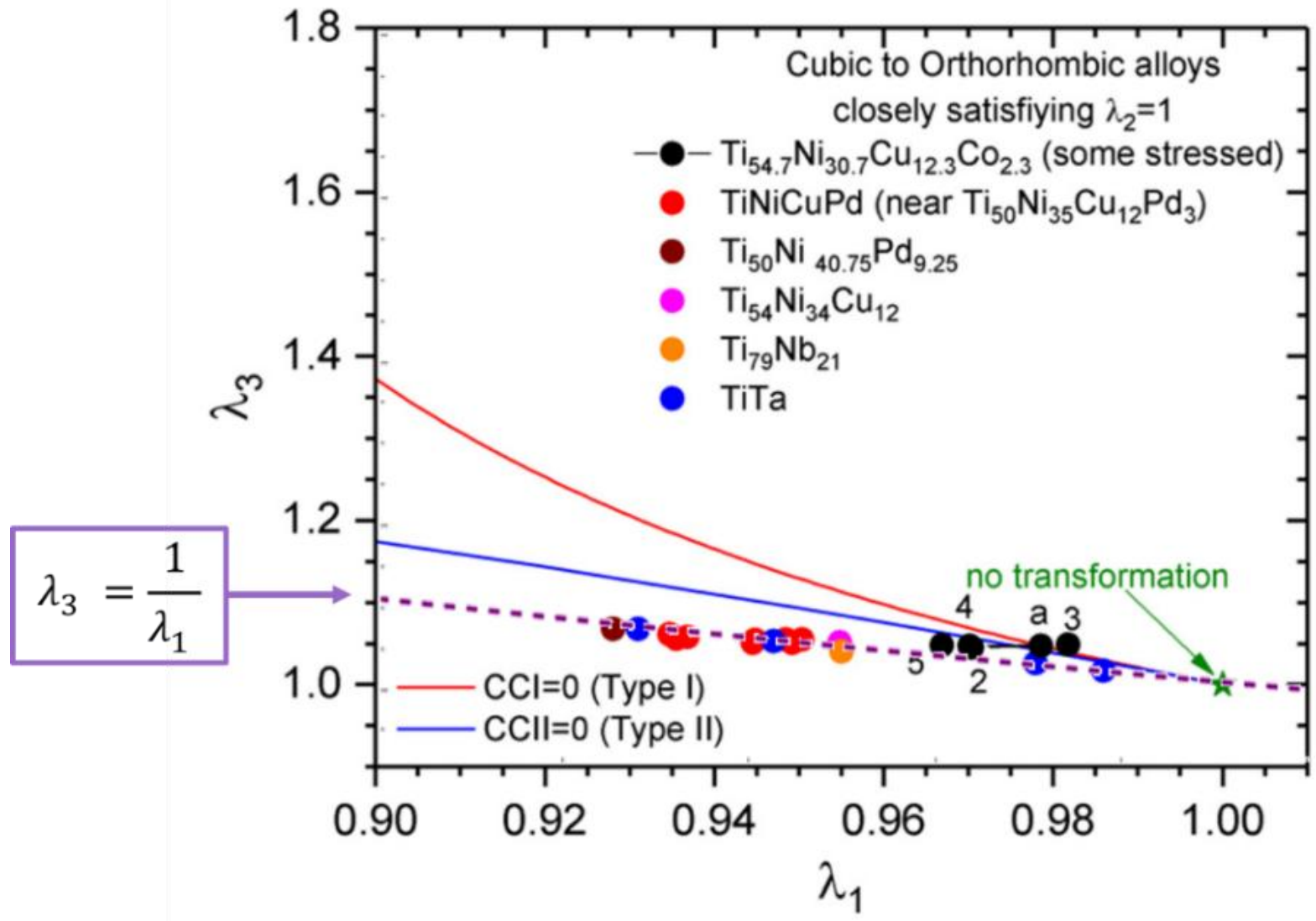

Figure 1. Plot $(\lambda_1, \lambda_3)$ for different cubic to orthorhombic transformations (colored points) such that the supercompatibilty condition SC1 is nearly satisfied $\lambda_2 \approx 1$. This figure is reproduced from Fig.3 of Ref Gu *et al.* [15]. The dashed purple curve corresponds to the equation $\lambda_3 = \frac{1}{\lambda_1}$ that is obtained when the dilatation part of the lattice distortion is $\delta \ll 1$.

## 2.2 Geometrical construction of supercompatible martensite

Let us consider a martensitic phase transformation for which the lattice distortion is an IPS, which also implies that $\lambda_2 = 1$. We note $\mathbf{m}$ the shear plane (a unit vector of the reciprocal space), and $\mathbf{d}$ the shear direction (a vector of the direct space). We consider two martensite variants, indexed 1 and 2, linked by a twin. The mirror plane of the twin is rational for a type I twin and irrational for type II twin. We call $\mathbf{F}_1$ and $\mathbf{F}_2$ their IPS lattice distortions given by:

$$\begin{aligned} \mathbf{F}_1 &= \mathbf{I} + \mathbf{d}_1\mathbf{m}_1^{\mathrm{t}} \\ \mathbf{F}_2 &= \mathbf{I} + \mathbf{d}_2\,\mathbf{m}_2^{\mathrm{t}} \end{aligned} \tag{8}$$

The interface between the two martensite variants is fully coherent if and only if $\mathbf{d}_1 = \mathbf{d}_2 = \mathbf{d}$. The norm of the shear vector depends on the pure shear amplitude $\tau$ and dilatation $\delta$ by $\|\mathbf{d}\| = \sqrt{\tau^2 + \delta^2}$.

The intersection of the shear planes associated with $\mathbf{F}_1$ and $\mathbf{F}_2$, noted $\hat{\mathbf{e}}$, is such that $\mathbf{F}_1\hat{\mathbf{e}} = \mathbf{F}_2\hat{\mathbf{e}} = \hat{\mathbf{e}}$. We consider $\hat{\mathbf{e}}^{\perp}$, the plane normal to $\hat{\mathbf{e}}$. Since the dilatation part of an IPS is normal to the shear plane, the dilatation vectors of $\mathbf{F}_1$ and $\mathbf{F}_2$ necessarily belong to $\hat{\mathbf{e}}^{\perp}$. The shear part however should be decomposed into a component $\tau_{\perp}$ in the plane $\hat{\mathbf{e}}^{\perp}$, and a component $\tau_{\parallel}$ along the vector $\hat{\mathbf{e}}$. Consequently, the shear directions of the IPS are written as $\mathbf{d} = \mathbf{d}_{\perp} + \mathbf{d}_{\parallel}$, with $\mathbf{d}_{\perp} = \tau_{\perp}(\mathbf{m} \times \hat{\mathbf{e}}) + \delta\, \mathbf{m}$ and $\mathbf{d}_{\parallel} = \tau_{\parallel}\hat{\mathbf{e}}$, each of them with their own index 1 or 2 omitted here for safe of clarity. The lattice distortions of the two variants in the plane $\hat{\mathbf{e}}^{\perp}$ are geometrically represented in Figure 2. In this figure, $\mathbf{p}$ the mirror plane between them was arbitrarily positioned vertically. Note that $\mathbf{p}$ refers to a plane (vector of the reciprocal space), and its normal is the direction $\mathbf{n}$ (vector of the direct space) that has the same meaning as in the SC equations (2) and (3). In an orthonormal basis, the two vectors are equal, but in a crystallographic basis, especially for non-cubic crystals, the two vectors should be distinguished. The twin plane $\mathbf{p}$ is directly inherited by correspondence from the austenite mirror plane for type I twins (see section 3). In this figure, the shear planes of the variants are oriented exactly such that the shear directions $\mathbf{d}_{\perp}$ of each variant come in coincidence and become equal. This coincidence is necessarily reached on the vertical mirror plane. With such a construction, the interface between the variant 1 and 2 is perfectly coherent.

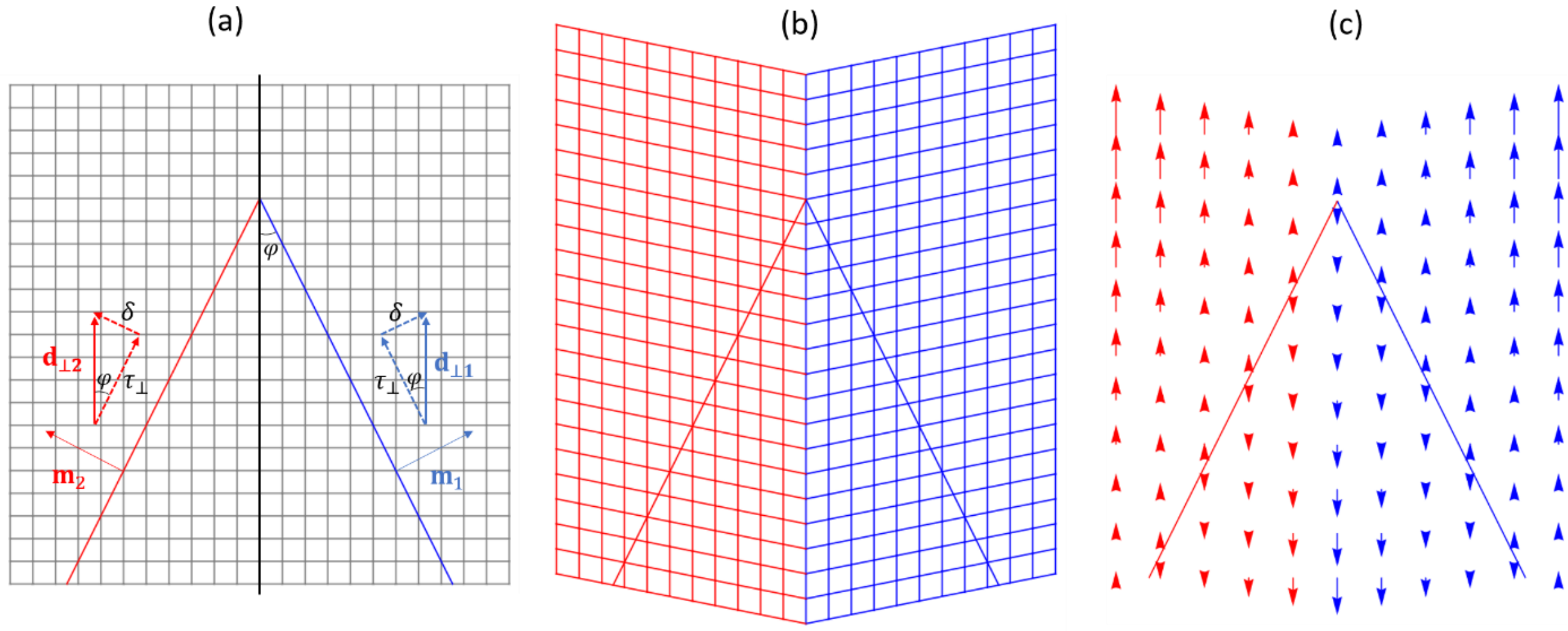

Figure 2. 2D representation in the plane $\hat{\mathbf{e}}^{\perp}$ of the lattice distortions of two IPS martensite variants. (a) Austenite phase that will be transformed into two twin-related variants 1 and 2. The planes $\mathbf{m}$ that will become the austenite/martensite coherent interface are the half-lines, in blue for variant 1 and in red for variant 2. The plane $\mathbf{p}$ that will become the interface plane between the variants is the black vertical line. (b) Same region after full martensitic transformation, i.e. after distortion $\mathbf{F}_1$ in the right half side (in blue), and $\mathbf{F}_2$ in the left half side (in red). (c) Displacement field between (a) and (b). For this figure, we used the following values: shear amplitude $\tau_{\perp} = 0.2$, dilatation $\delta = 0.1$.

We note $m_x, m_y$ the coordinates of $\mathbf{m}_1$ the shear plane of variant 1 written in the orthonormal basis $(\hat{\boldsymbol{x}}, \hat{\boldsymbol{y}})$, where $\boldsymbol{x}$ and $\boldsymbol{y}$ are the horizontal and vertical directions of Figure 2. The shear planes and shear directions of the variants 1 and 2 written in this basis are:

$$
\begin{aligned}
&\mathbf{m}_1 = (m_x, m_y) \quad \text{and} \quad \mathbf{d}_{\perp 1} = \tau_{\perp} \begin{bmatrix} -m_y \\ m_x \end{bmatrix} + \delta \begin{bmatrix} m_x \\ m_y \end{bmatrix} \\
&\mathbf{m}_2 = (-m_x, m_y) \text{ and } \mathbf{d}_{\perp 2} = \tau_{\perp} \begin{bmatrix} m_y \\ m_x \end{bmatrix} + \delta \begin{bmatrix} -m_x \\ m_y \end{bmatrix}
\end{aligned} \tag{9}
$$

The (A/M) IPS compatibility and the (M/M) twin compatibility are compatible together if the shear vectors $\mathbf{d}$ are equal and located in the vertical mirror plane, which imposes that $\mathbf{d}_{\parallel 1} = \mathbf{d}_{\parallel 2}$ and $\mathbf{d}_{\perp 1} = \mathbf{d}_{\perp 2}$, and that their $\boldsymbol{x}$-coordinate is null. These conditions are verified when

$$
m_x = \frac{\tau_{\perp}}{\sqrt{\tau_{\perp}^2 + \delta^2}}, \; m_y = \frac{\delta}{\sqrt{\tau_{\perp}^2 + \delta^2}} \tag{10}
$$

By construction, the shear vector in the basis $(\hat{\boldsymbol{x}}, \hat{\boldsymbol{y}})$ is $\mathbf{d}_{\perp 1} = \mathbf{d}_{\perp 2} = \begin{bmatrix} 0 \\ d \end{bmatrix}$ with $\|\mathbf{d}_{\perp}\| = \sqrt{\tau_{\perp}^2 + \delta^2}$ . The shear planes of the variants 1 and 2 makes an angle $\phi$ with the vertical mirror plane given by

$$\tan(\phi) = \frac{\delta}{\tau_{\perp}} \tag{11}$$

Note that if the martensite transformation is a simple shear, $\delta = 0$, thus $\phi = 0$, the shear planes of the variants 1 and 2 come in coincidence with the mirror plane between them. If the martensite transformation is a pure dilatation, $\tau = 0$, thus $\phi = 90°$, the shear planes of the variants 1 and 2 become also a unique plane perpendicular to the mirror plane.

Equation (11) gives the angle between the habit plane of the individual martensite variants and the junction (mirror) plane between them. Another formula can be found between $\phi$, $\mathbf{d}$, the shear direction of the IPS, and $\mathbf{a}$, the shear direction of the twin. Since $\mathbf{F}_1 = \mathbf{I} + \mathbf{d}\,\mathbf{m}_1^{\mathrm{t}}$ and $\mathbf{F}_2 = \mathbf{I} + \mathbf{d}\,\mathbf{m}_2^{\mathrm{t}}$, the twin shear matrix between the variants 1 and 2 is given in austenite crystallographic basis by $\mathbf{F}_1\mathbf{F}_2^{-1}$. Using equation (5), we obtain $\mathbf{F}_1\,\mathbf{F}_2^{-1} = \mathbf{I} + \mathbf{d}\,(\mathbf{m}_1^{\mathrm{t}} - \frac{1}{1+\delta}\mathbf{m}_2^{\mathrm{t}}) - \frac{1}{1+\delta}\mathbf{d}\,\mathbf{m}_1^{\mathrm{t}}\,\mathbf{d}\,\mathbf{m}_2^{\mathrm{t}}$, with $\mathbf{m}_1^{\mathrm{t}}\,\mathbf{d} = \delta$.Thus, $\mathbf{F}_1\mathbf{F}_2^{-1} = \mathbf{I} + \mathbf{d}\,(\mathbf{m}_1^{\mathrm{t}} - \frac{1}{1+\delta}\mathbf{m}_2^{\mathrm{t}} - \frac{\delta}{1+\delta}\mathbf{m}_2^{\mathrm{t}}) = \mathbf{I} + \mathbf{d}\,(\mathbf{m}_1^{\mathrm{t}} - \mathbf{m}_2^{\mathrm{t}}) = \mathbf{I} + 2\cos\varphi\,\mathbf{d}\,\mathbf{p}^{\mathrm{t}}$, where $\mathbf{p}$ is the mirror plane of the twin, and its normal is the direction $\mathbf{n}$ that has the same meaning as in the SC equations (2) and (3). Therefore, $\mathbf{d}$ and $\mathbf{a}$ are linked by the "shear/shear" equation

$$\mathbf{a} = 2\cos(\varphi)\,\mathbf{d} \tag{12}$$

where $\cos(\varphi)$ is the scalar product of the unit vectors between the shear plane of the individual variants and the mirror plane of the twin between them, i.e. $\cos(\varphi) = \mathbf{m}_1^{\mathrm{t}}\,\mathbf{n} = -\mathbf{m}_2^{\mathrm{t}}\,\mathbf{n}$.

The lattice distortions of two twin-related variants formed inside the surrounding austenite matrix are illustrated in Figure 3. It can be checked in this figure that both the A/M interfaces and the M/M junction plane are coherent.

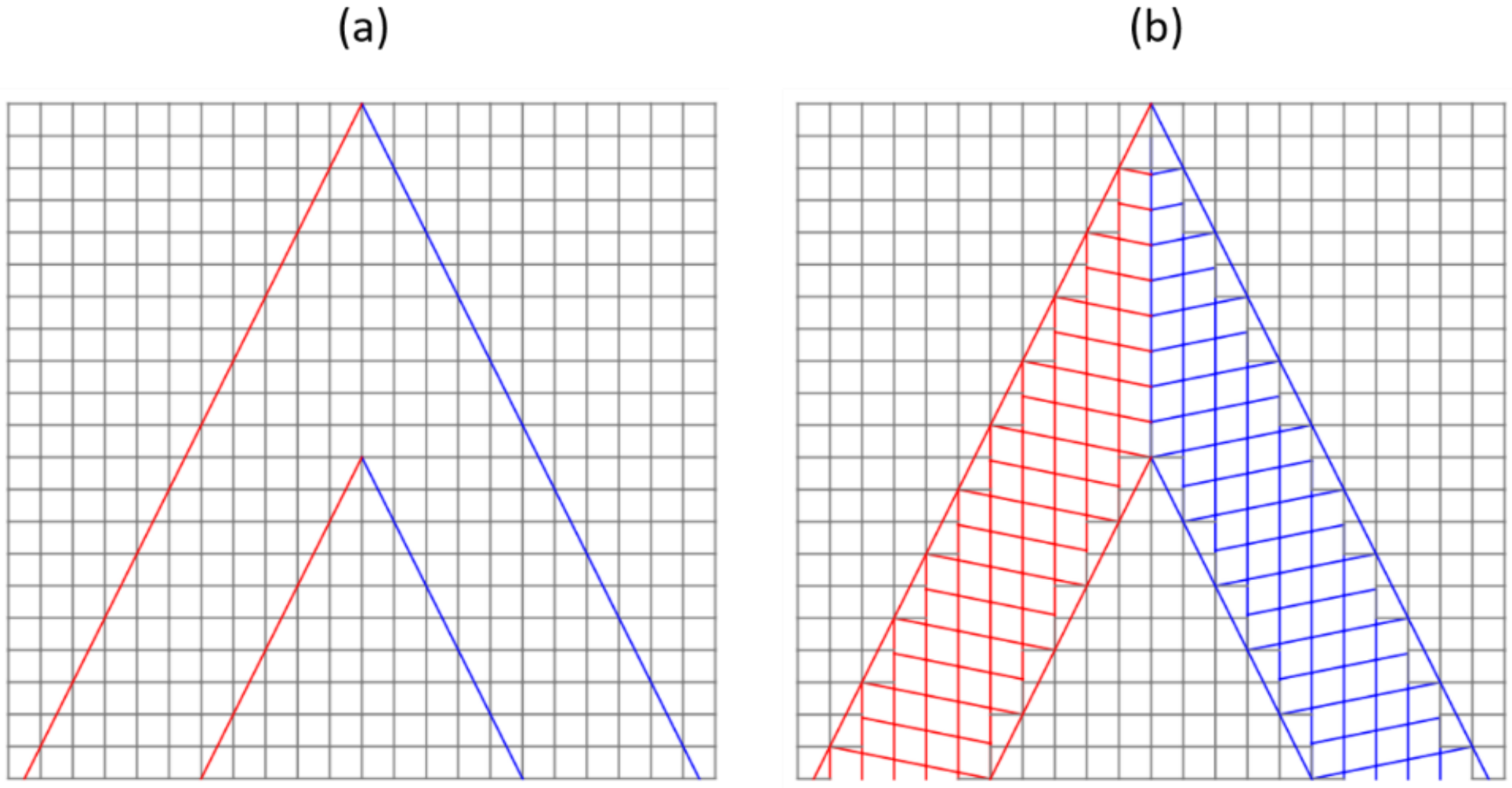

Figure 3. Two martensite variants (1 blue lattice, 2 red lattice) formed as plates inside a surrounding austenite matrix (grey square lattice). (a) Before the transformation. (b) After transformation. The distortion parameters are $\tau_{\perp} = 0.2$, $\delta = 0.1$

Now, we consider a heterogeneous laminate martensite product constituted of the variants 1 and 2 in volume fractions $f$ and $1 - f$, respectively. The average lattice distortion $\mathbf{P}$ is an IPS whatever $f$ because the shear vector is common to the two variants. Indeed,

$$\mathbf{P} = f\ \mathbf{F}_1 + (1-f)\ \mathbf{F}_2 = \mathbf{I} + \mathbf{d} \otimes (f\ \mathbf{m}_A + (1-f)\ \mathbf{m}_B) \tag{13}$$

The habit plane of the laminate martensite product is the plane

$$\mathbf{m} = f\ \mathbf{m}_A + (1-f)\ \mathbf{m}_B \tag{14}$$

Some examples are shown in Figure 4. In the specific case where $f = 1 - f = 1/2$, the normal $\mathbf{m}$ is perpendicular to the twin plane, as illustrated in Figure 4b.

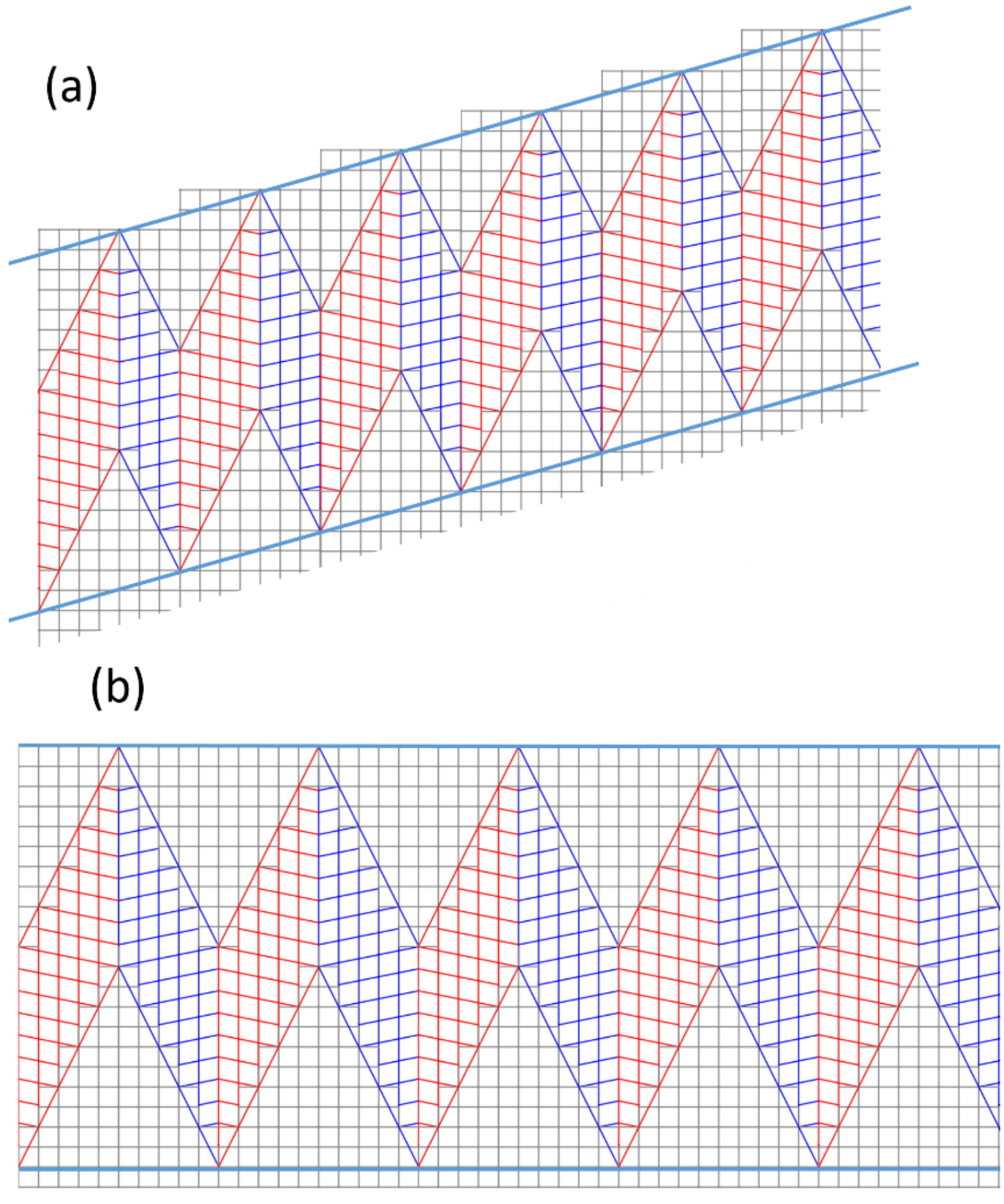

Figure 4. Some examples of different habit planes of laminate structure made of two martensite variants in different proportions. a) Habit plane formed with $f = 4/5$, $1 - f = 3/7$. (b) Habit plane formed with $f = 1 - f = 1/2$. Case $\tau_{\perp} = 0.2$, $\delta = 0.1$, vertical twin mirror plane.

We have shown that when the lattice distortion of a martensitic transformation is an IPS (A/M compatibility), and when two martensite variants are twin related (M/M compatibility), an orientation of the shear planes of the individual variants relatively to the twin mirror plane given by equation (11) allows for a perfect compatibility of the A/M/M system, and bi-variant laminate products can be formed in any volume fraction $f$ with an habit plane given by equation (14), i.e. supercompatibility is obtained.

## 2.3 Checking the SC2 and SC3 conditions

The geometric construction of supercompatible martensite detailed in the previous section should verify the SC equations. Since the lattice distortion is an IPS, SC1 is immediately verified.

We consider the condition SC2. In the eigen basis $(\mathbf{e}_1, \mathbf{e}_2, \mathbf{e}_3)$, $\mathbf{U} = \mathrm{diag}\,(\lambda_1, 1, \lambda_3)$, and $\mathbf{U}^2 - \mathbf{I} = \mathrm{diag}\,(\lambda_1^2 - 1, 0,\ \lambda_3^2 - 1)$. All the components of $\mathrm{cof}\,(\mathbf{U}^2 - \mathbf{I})$ are null except the component (2,2) that is equal to $(\lambda_1^2 - 1)(\lambda_3^2 - 1)$. Thus, $\mathbf{U}\ \mathrm{cof}(\mathbf{U}^2 - \mathbf{I})$ is also a matrix for which all the components are null except in (2,2). Consequently, $\mathbf{U}\ \mathrm{cof}(\mathbf{U}^2 - \mathbf{I})\ \mathbf{n}$ is a direction parallel to the direction $\mathbf{e}_2$. Since $\mathbf{e}_2$ is perpendicular to $\mathbf{d}$, and $\mathbf{d}$ is parallel to $\mathbf{a}$ by equation (11), $\mathbf{a}\ \mathbf{U}\ \mathrm{cof}(\mathbf{U}^2 - \mathbf{I})\ \mathbf{n} = 0$. Consequently SC2 = 0.

We consider the condition SC3. From equation (12), it appears that the square of the norm of $\mathbf{a}$ is $\mathbf{a}^2 = 4\cos^2(\phi)\,(\tau^2 + \delta^2)$. By definition $\mathbf{p}^2 = 1$. Direct calculation of equation (3) with equations (6) and (7) shows that $\mathrm{SC3} = \sin^2(\phi)\,\tau^2 - \cos^2(\phi)\,\delta^2$. Since $\tau^2 = \tau_\perp^2 + \tau_\parallel^2 \geq \tau_\perp^2$, we obtain that SC3 $\geq \sin^2(\phi)\ \tau_\perp^2 - \cos^2(\phi)\,\delta^2$. Using equation (11), we get SC3 $\geq \cos^2(\phi)\,\delta^2 - \cos^2(\phi)\,\delta^2 = 0$.

This section showed that supercompatibility is obtained when a) the A/M interface is invariant, i.e the lattice distortion is an IPS, b) the martensite variants are linked by a M/M twin relationship, and c) the shear/shear relationship (12) exists between the shear direction of the IPS and the shear direction of the twin. These conditions are those of a full compatibility of the A/M/M system, i.e. supercompatibility. We will propose in section 4 an approach to deduce the supercompatibility equations from these criteria. Since the method uses the same crystallographic tools as those of the CT [19], a short reminder on the CT and its main equations is necessary.

# 3 The correspondence theory in brief

## 3.1 The metric tensors and their central role in crystallography

For any structure, the conventional crystallographic basis $\boldsymbol{\mathcal{B}}_c = (\mathbf{a}, \mathbf{b}, \mathbf{c})$ has an associated its metric tensor given by

$$\boldsymbol{\mathcal{M}} = \begin{pmatrix} \mathbf{a}^2 & \mathbf{b}^{\mathrm{t}}\,\mathbf{a} & \mathbf{c}^{\mathrm{t}}\,\mathbf{a} \\ \mathbf{a}^{\mathrm{t}}\,\mathbf{b} & \mathbf{b}^2 & \mathbf{c}^{\mathrm{t}}\,\mathbf{b} \\ \mathbf{a}^{\mathrm{t}}\,\mathbf{c} & \mathbf{b}^{\mathrm{t}}\,\mathbf{c} & \mathbf{c}^2 \end{pmatrix} = \begin{pmatrix} \|\mathbf{a}\|^2 & \|\mathbf{b}\|\|\mathbf{a}\|\,cos(\gamma) & \|\mathbf{c}\|\|\mathbf{a}\|\,cos(\beta) \\ \|\mathbf{a}\|\|\mathbf{b}\|\,cos(\gamma) & \|\mathbf{b}\|^2 & \|\mathbf{c}\|\|\mathbf{b}\|\,cos(\alpha) \\ \|\mathbf{a}\|\|\mathbf{c}\|\,cos(\beta) & \|\mathbf{b}\|\|\mathbf{c}\|\,cos(\alpha) & \|\mathbf{c}\|^2 \end{pmatrix} \tag{15}$$

where $\|\mathbf{a}\|$, $\|\mathbf{b}\|$, $\|\mathbf{c}\|$ are the lengths of the vectors, and $\alpha, \beta, \gamma$ are the angles between them, i.e. $\alpha = (\widehat{\mathbf{b}, \mathbf{c}})$, $\beta = (\widehat{\mathbf{a}, \mathbf{c}})$, $\gamma = (\widehat{\mathbf{a}, \mathbf{b}})$. Note that the determination of the metric tensor does not require introducing any orthonormal basis; it just assumes the existence of measures for distances (a ruler) and angles (a protractor). The metric tensor is nothing else than the coordinate transformation matrix from the reciprocal space to the direct space, and can thus be noted $\boldsymbol{\mathcal{M}} = [\boldsymbol{\mathcal{B}}_c^* \rightarrow \boldsymbol{\mathcal{B}}_c]$. It is symmetric, $\boldsymbol{\mathcal{M}} = \boldsymbol{\mathcal{M}}^{\mathrm{t}}$. In addition, $\boldsymbol{\mathcal{M}}^* = [\boldsymbol{\mathcal{B}}_c \rightarrow \boldsymbol{\mathcal{B}}_c^*] = \boldsymbol{\mathcal{M}}^{-1}$. The scalar product between two vectors $\mathbf{u}$ and $\mathbf{v}$ of the direct space is determined by expressing one vector into the reciprocal space thanks to the metric tensor: $(\mathbf{u} \cdot \mathbf{v}) = \mathbf{u}^{\mathrm{t}}\ \boldsymbol{\mathcal{M}}\ \mathbf{v}$. The norm $\|\mathbf{u}\|$ of a vector $\mathbf{u}$ of the direct space, and the norm $\|\mathbf{p}\|^*$ of a vector $\mathbf{p}$ of the reciprocal space are given by $\|\mathbf{u}\| = \sqrt{\mathbf{u}^{\mathrm{t}}\ \boldsymbol{\mathcal{M}}\ \mathbf{u}}$ and $\|\mathbf{p}\|^* = \sqrt{\mathbf{p}^{\mathrm{t}}\ \boldsymbol{\mathcal{M}}^*\ \mathbf{p}}$, respectively. The notation $\tilde{\mathbf{u}}$ applied to a direct vector means that $\mathbf{u}$ is normalized by $\|\mathbf{u}\|$, i.e. $\tilde{\mathbf{u}} = \frac{\mathbf{u}}{\|\mathbf{u}\|^*}$, and the notation $\tilde{\mathbf{p}}$ applied to a reciprocal vector means that $\mathbf{p}$ is normalized by $\|\mathbf{p}\|^*$, i.e. $\tilde{\mathbf{p}} = \frac{\mathbf{p}}{\|\mathbf{p}\|^*}$. The inter-reticular distance $d_{hkl}$ between the layers of a plane $\mathbf{p}$ of Miller indices $\mathbf{p} = (h\ k\ l)$ is $d_{hkl} = \frac{1}{\|\mathbf{p}\|^*}$ . The unit direction normal to a plane $\mathbf{p}$ is a vector of the direct space $\tilde{\mathbf{n}}$ given by $\tilde{\mathbf{n}} = \boldsymbol{\mathcal{M}}^*\ \tilde{\mathbf{p}}$. It can be verified that $\tilde{\mathbf{n}}^{\mathrm{t}} \boldsymbol{\mathcal{M}}\ \tilde{\mathbf{n}} = \tilde{\mathbf{n}}^{\mathrm{t}}\ \tilde{\mathbf{p}} = 1$. For sake of simplicity, the symbol "~" is not used when vectors are explicitly unit vectors.

Despite its central role in crystallography, the metric tensor has always been ignored in the PTMC, from its early beginning 70 years ago [1,2] up to its modern versions [4,9]. The PTMC procedures are mainly based on polar decomposition derived from continuum mechanics, which requires the help of an orthonormal basis. When working with a parent cubic phase, the orthonormal basis can be confused with the crystallographic basis because it differs from it only by a proportionality factor that is the square of the lattice parameter. In an orthonormal basis, any matrix $\mathbf{F}$ can be decomposed $\mathbf{F} = \mathbf{R}\,\mathbf{U}$, where the symmetric matrix $\mathbf{U}$ is calculated from $\mathbf{F}^{\mathrm{t}}\,\mathbf{F} = \mathbf{U}^{\mathrm{t}}\,\mathbf{U}$, but the method works only because $\mathbf{R}^{\mathrm{t}}\,\mathbf{R} = \mathbf{I}$ since a rotation is an isometry written in an orthonormal basis. However, the last equality is generally wrong if the basis is not orthonormal. In the general case, only the equation of conservation of metric by an isometry holds. For any isometry (rotation or reflection) of matrix $\mathbf{g}$, and any pair of vectors $\mathbf{u}$ and $\mathbf{v}$, the scalar product $(\mathbf{g}\,\mathbf{u} \cdot \boldsymbol{g}\,\mathbf{v}) = \mathbf{u}^{\mathrm{t}}\,\mathbf{g}^{\mathrm{t}}\,\boldsymbol{\mathcal{M}}\,\mathbf{g}\,\mathbf{v} = (\mathbf{u} \cdot \mathbf{v}) = \mathbf{u}^{\mathrm{t}}\,\boldsymbol{\mathcal{M}}\,\mathbf{v}$, which leads to

$$\mathbf{g}^{\mathrm{t}}\,\boldsymbol{\mathcal{M}}\,\mathbf{g} = \boldsymbol{\mathcal{M}} \tag{16}$$

but not necessarily that $\mathbf{g}^{\mathrm{t}}\,\mathbf{g} = \mathbf{I}$. The failure of polar decomposition for non- cubic crystals can be shown with a simple 2D hexagonal lattice of crystallographic basis $(\mathbf{a}, \mathbf{b})$, with $\|\mathbf{a}\| = \|\mathbf{b}\| = 1$. The metric tensor is $\boldsymbol{\mathcal{M}} = \begin{pmatrix} 1 & -1/2 \\ -1/2 & 1 \end{pmatrix}$. In the basis $(\mathbf{a}, \mathbf{b})$, the symmetry rotation matrix of + 60° is simply $\mathbf{R}^{+} = \begin{pmatrix} 1 & -1 \\ 1 & 0 \end{pmatrix}$ and its inverse is $\mathbf{R}^{-} = \begin{pmatrix} 0 & -1 \\ -1 & 0 \end{pmatrix}$. It can be checked that $(\mathbf{R}^{+})^{\mathrm{t}}\,\boldsymbol{\mathcal{M}}\,\mathbf{R}^{+} = (\mathbf{R}^{-})^{\mathrm{t}}\,\boldsymbol{\mathcal{M}}\,\mathbf{R}^{-} = \boldsymbol{\mathcal{M}}$. It is however clear that $\mathbf{R}^{-}$ is not the transpose of $\mathbf{R}^{+}$. To our opinion, polar decomposition and stretch matrices are not adapted to crystallographic problems implying non-cubic phases.

## 3.2 The three types of transformation matrices

The crystallography of martensitic phase transformations can be described with three types of matrix, each of them covering a different and complementary aspect of the transformation: lattice distortion, orientation relationship (OR), and correspondence [20]. Let us explain them for a A → M martensitic transformation. The lattice distortion takes the form of an active matrix $\mathbf{F}^{\mathrm{A}}$. Any austenite direction $\mathbf{u}^{\mathrm{A}}$ is transformed by the distortion into a new direction $\mathbf{u}^{\mathrm{A}'} = \mathbf{F}^{\mathrm{A}}\,\mathbf{u}^{\mathrm{A}}$ in the same basis. The distortion matrix $\mathbf{F}^{\mathrm{A}}$ is usually written in the conventional crystallographic basis of the parent phase $\boldsymbol{\mathcal{B}}_c^{\mathrm{A}} = (\mathbf{a}^{\mathrm{A}}, \mathbf{b}^{\mathrm{A}}, \mathbf{c}^{\mathrm{A}})$ by $\mathbf{F}_c^{\mathrm{A}} = [\boldsymbol{\mathcal{B}}_c^{\mathrm{A}} \to \boldsymbol{\mathcal{B}}_c^{\mathrm{A}'}] = (\mathbf{a}^{\mathrm{A}'}, \mathbf{b}^{\mathrm{A}'}, \mathbf{c}^{\mathrm{A}'})$, writing in columns the coordinates in $\boldsymbol{\mathcal{B}}_c^{\mathrm{A}}$ of $\mathbf{a}^{\mathrm{A}'}, \mathbf{b}^{\mathrm{A}'}, \mathbf{c}^{\mathrm{A}'}$ that are the images by distortion of the austenite basis vectors. The OR between the austenite crystal and a one of the martensite variants is given by the coordinate transformation matrix $\mathbf{T}^{\mathrm{A}\to\mathrm{M}} = [\boldsymbol{\mathcal{B}}_c^{\mathrm{A}} \to \boldsymbol{\mathcal{B}}_c^{\mathrm{M}}]$. It is passive as it changes the coordinates between the parent and daughter bases of any fixed vector $\mathbf{u}$ as follows $\mathbf{u}_{/\mathrm{A}} = \mathbf{T}^{\mathrm{A}\to\mathrm{M}}\,\mathbf{u}_{/\mathrm{M}}$. The correspondence matrix $\mathbf{C}^{\mathrm{M}\to\mathrm{A}}$ gives in columns the coordinates of $\mathbf{a}^{\mathrm{A}'}, \mathbf{b}^{\mathrm{A}'}, \mathbf{c}^{\mathrm{A}'}$ in the martensite basis $\boldsymbol{\mathcal{B}}_c^{\mathrm{M}}$. Explicitly, $\mathbf{C}^{\mathrm{M}\to\mathrm{A}} = \left(\mathbf{a}^{\mathrm{A}'}_{/\mathrm{M}}, \mathbf{b}^{\mathrm{A}'}_{/\mathrm{M}}, \mathbf{c}^{\mathrm{A}'}_{/\mathrm{M}}\right)$. Any direction $\mathbf{u}$ becomes after lattice distortion a direction $\mathbf{u}'$ that is written in the martensite crystallographic basis by $\mathbf{u}'_{/\mathrm{M}} = \mathbf{C}^{\mathrm{M}\to\mathrm{A}}\,\mathbf{u}_{/\mathrm{A}}$, where $\mathbf{u}_{/\mathrm{A}}$ is the vector $\mathbf{u}$ written in $\boldsymbol{\mathcal{B}}_c^{\mathrm{A}}$. This equation will be often written $\mathbf{u}_{\mathrm{M}} = \mathbf{C}^{\mathrm{M}\to\mathrm{A}}\,\mathbf{u}_{\mathrm{A}}$ for simplicity and to emphasis that the crystallographic direction of austenite becomes by transformation a crystallographic direction of martensite. The components of the correspondence matrix are integers, or half-integers when face-centred or body centred Bravais lattices are involved. It can be also shown that $\mathbf{T}^{\mathrm{A}\to\mathrm{M}} = \left(\mathbf{T}^{\mathrm{M}\to\mathrm{A}}\right)^{-1}$ and $\mathbf{C}^{\mathrm{A}\to\mathrm{M}} = \left(\mathbf{C}^{\mathrm{M}\to\mathrm{A}}\right)^{-1}$. The three transformation matrices are linked by the equation $\mathbf{C}^{\mathrm{M}\to\mathrm{A}} = \mathbf{T}^{\mathrm{M}\to\mathrm{A}}\,\mathbf{F}^{\mathrm{A}}$.

## 3.3 The transformation twins from austenite symmetries

The correspondence theory assumes the existence of a natural OR between austenite and martensite, but this OR is not strict and deviations are allowed to get compatibility between the variants. The CT uses the fact that the compatibility is obtained locally between two variants by a symmetry the parent phase.

If the parent symmetry is a reflection, the rational mirror plane becomes *by correspondence* the mirror plane between the martensite variants, and the transformation twin between them is type I. If the parent symmetry is a two-fold rotation, the rational 180° rotation axis becomes *by correspondence* the 180° rotation axis between the martensite variants, and the transformation twin between them is type II. If the parent symmetry is a $n$-fold rotation with $n \neq 2$ (for example $n$ = 3, 4 or 6, depending on the point group of the parent phase), the rational rotation axis becomes the $n$-fold rotation axis between the martensite variants, and for such cases, no plane (rational or irrational) can be perfectly compatible, however, some planes called "weak planes" show minimal intrinsic distortion and can thus take the role of junction plane between the variants [21]. Each pair of twin-related variants implies the existence of an OR that slightly deviate from the natural OR. These new ORs are called "closing gap" ORs in the CT [19]. Contrarily to the PTMC, the CT allows for direct calculation of the twins without using stretch matrices and polar decompositions. Let us recall here its main equations. The correspondence matrix $\mathbf{C}^{\mathrm{M}\to\mathrm{A}}$ is given relatively to the variant $\mathrm{M}_1 = \mathrm{M}$.

We consider two variants $\mathrm{M}_1$ and $\mathrm{M}_2$ joined by a reflection symmetry of austenite on the plane $\mathbf{p}_\mathrm{A}$. In the case of cubic austenite, $\mathbf{p}_\mathrm{A} = \{100\}\ or\ \{110\}$. The reflection matrix is noted $m^\mathrm{A}$. The plane $\mathbf{p}_\mathrm{A}$ is transformed by correspondence into the martensitic plane $\mathbf{p}_\mathrm{M}$ that has the same Millers indices in $\mathrm{M}_1$ and $\mathrm{M}_2$, and the parallelism between $\mathbf{p}_\mathrm{A}$ and $\mathbf{p}_\mathrm{M}$ is maintained thanks to the closing-gap rotation. The two martensite variants are related by a type I twin for which the martensite invariant plane $\mathbf{p}_\mathrm{M}$ (often noted $\mathrm{K}_1$) is directly deduced from $\mathbf{p}_\mathrm{A}$ by

$$\mathbf{p}_\mathrm{M} = \left(\mathbf{C}^{\mathrm{M}\to\mathrm{A}}\right)^{-\mathrm{t}} \mathbf{p}_\mathrm{A} \qquad (17)$$

This plane is rational because the correspondence matrix is rational and $\mathbf{p}_\mathrm{A}$ is rational. This plane is also the junction plane between the martensite variants. The "shear" amplitude, $s$, and the "shear" direction of the transformation twin between $\mathrm{M}_1$ and $\mathrm{M}_2$, $\mathbf{a}_\mathrm{M}$ (often noted $\eta_1$), depend uniquely on $\mathcal{M}_\mathrm{M}$, the metric of martensite, and not on the metric of austenite. They are given by equations derived from Bevis and Crocker's work on deformation twinning [22]:

$$s^2 = \mathrm{tr}\left(\mathbf{C}_{\mathrm{int}}^{\mathrm{t}}\, \mathcal{M}_\mathrm{M}\, \mathbf{C}_{\mathrm{int}}\, \mathcal{M}_\mathrm{M}^{-1}\right) - 3 \qquad (18)$$

$$\mathbf{a}_\mathrm{M} = -(\mathbf{C}_{\mathrm{int}} + \mathbf{I})\, \mathbf{n}_\mathrm{M} \qquad (19)$$

where $\mathbf{C}_{\mathrm{int}} = \mathbf{C}^{\mathrm{M}_1\to\mathrm{M}_2} = \mathbf{C}^{\mathrm{M}\to\mathrm{A}}\, m^\mathrm{A}\, \mathbf{C}^{\mathrm{A}\to\mathrm{M}}$ is the intercorrespondence matrix between $\mathrm{M}_1$ and $\mathrm{M}_2$. In equation (19), $\mathbf{n}_\mathrm{M}$ is the normal to the martensite mirror plane, i.e. $\mathbf{n}_\mathrm{M} = \mathcal{M}_\mathrm{M}^{-1}\, \mathbf{p}_\mathrm{M}$. The ("shear" plane, "shear" direction) system of the transformation twin is $(\mathbf{p}_\mathrm{M}, \mathbf{d}_\mathrm{M})$. Note that transformation twins should be however distinguished from the deformation twins because the variants form together during cooling, and the twin relation between them does not imply that one variant is mechanically transformed into another. Even in the case of "detwinning", i.e. variant reorientation under strain, we do not believe that the variant $\mathrm{M}_1$ is directly transformed into $\mathrm{M}_2$ by a simple shear because the atoms would interpenetrate too much; we think that trajectories close to a double path $\mathrm{M}_1 \to \mathrm{A} \to \mathrm{M}_2$ would be more realistic. The transformation twin imposes a local closing-gap OR in which the austenite and martensite "shear" planes and directions are parallel:

$$\begin{cases} \mathbf{p}_\mathrm{A} \parallel \mathbf{p}_\mathrm{M} \text{ (rational "shear" plane } \mathrm{K}_1) \\ \mathbf{a}_\mathrm{A} \parallel \mathbf{a}_\mathrm{M} \text{ (irrational "shear" direction } \eta_1) \end{cases} \qquad (20)$$

Note that $\mathbf{a}_\mathrm{A}$ is the austenite "shear" direction of the transformation twin obtained from the martensite shear direction $\mathbf{a}_\mathrm{M}$ by $\mathbf{a}_\mathrm{A} = \mathbf{C}^{\mathrm{A}\to\mathrm{M}}\, \mathbf{a}_\mathrm{M}$; it has the same meaning as the vector **a** in the SC2 and SC2 equations (2) and (3), respectively, at least if the austenite crystal is cubic. The "shear" plane of the twin $\mathbf{p}_\mathrm{A}$ has for normal the vector **n** in the SC equations.

We consider now two variants $M_1$ and $M_2$ joined by a 180° rotation symmetry of austenite $R_\pi^A$. We note $\mathbf{a}_A$ the austenite rotation axis of $R_\pi^A$. In the case of cubic austenite, $\mathbf{a}_A = \langle 100 \rangle$ *or* $\langle 110 \rangle$. It is transformed by correspondence into the martensite direction $\mathbf{a}_M$. The parallelism between $\mathbf{a}_A$ and $\mathbf{a}_M$ is maintained thanks to the close-gap OR. The two martensite variants are related by a type II twins for which the invariant direction $\mathbf{a}_M$ (often noted $\eta_2$) is directly deduced from $\mathbf{a}_A$ by

$$\mathbf{a}_M = \mathbf{C}^{M \to A} \, \mathbf{a}_A \tag{21}$$

This axis is rational because the correspondence matrix is rational and $\mathbf{a}_A$ is rational. It is the "shear direction" of the twin and is contained in the shear plane between the martensite variants. The "shear" plane is also the junction plane of the twin. The "shear" amplitude $s^*$ and the shear plane $\mathbf{jp}_M$ (often noted $K_2$) depend on the metric of the martensite $\boldsymbol{\mathcal{M}}_M$, and not on that of the austenite. They are

$$s^{*2} = \mathrm{tr}\left(\mathbf{C}_{int} \; \boldsymbol{\mathcal{M}}_M^{-1} \; \mathbf{C}_{int}^t \, \boldsymbol{\mathcal{M}}_M\right) - 3 \tag{22}$$

$$\mathbf{jp}_M = -(\mathbf{C}_{int}^* - \mathbf{I}) \; \mathbf{p}_M \tag{23}$$

where $\mathbf{C}_{int} = \mathbf{C}^{M_1 \to M_2} = \mathbf{C}^{M \to A} \, R_\pi^A \, \mathbf{C}^{A \to M}$. In equation (23), $\mathbf{p}_M$ is the plane normal to the shear direction $\mathbf{a}_M$, i.e. $\mathbf{p}_M = \boldsymbol{\mathcal{M}}_M \, \mathbf{a}_M$. The ("shear" plane, "shear" direction) system of the transformation twin is $(\mathbf{jp}_M, \; \mathbf{a}_M)$. The twin imposes a local closing-gap OR :

$$\begin{cases} \mathbf{a}_A \parallel \; \mathbf{a}_M \text{ (rational "shear" direction } \eta_2) \\ \mathbf{jp}_A \parallel \; \mathbf{jp}_M \text{ (irrational "shear" plane } K_2) \end{cases} \tag{24}$$

Note that $\mathbf{a}_A$ is the "shear" direction of the transformation twin; it has the same meaning as the vector **a** in the SC2 and SC2 equations (2) and (3), respectively. The "shear" plane of the twin $\mathbf{jp}_A$is deduced from $\mathbf{jp}_M$ by $\mathbf{jp}_A = \left(\mathbf{C}^{A \to M}\right)^{-1} \mathbf{jp}_M$; it has for normal the vector **n** in the equations (2) and (3).

We have shown in this section that the transformation twins can be calculated directly from the symmetries of the parent phase. Contrarily to the PTMC, the equations are simple and direct. They have also the advantages to show immediately the respective roles of symmetries and metrics. For example, equations (17) and (21) show that the mirror plane for type I twins and the 180° rotation axis for the type II twins depend only the correspondence, and not on the metrics, which explains why these twin elements are generic, i.e. insensitive to any change of lattice parameters. They also show that the other twin elements, i.e. the "shear" direction for the type I twins and the "shear" plane for the type II twins are non-generic because they depend on the metric of the daughter phase. The fact they do not depend on the metric of the parent phase is clear in the CT, but far from obvious in the PTMC because the stretch matrix mixes up the metrics of the parent and daughter phases (see section 6). The CT has also a more rigorous and efficient treatment of the symmetries to calculate the different types of twins, as explained in the following sections.

## 3.4 The correspondence variants

The PTMC explores all the possible pairs of stretch variants to check those for which the rank-1 conditions can be solved. This method implies many redundant and unnecessary calculations. In the CT, we use group theory to significantly reduce the number of calculations and keep a trace of the parent symmetry elements that create the transformation twins, i.e. the plane $\mathbf{p}_A$ for type I twins, and direction $\mathbf{a}_A$ for type II twins. For each of the three types of transformation matrices (distortion, orientation, correspondence), the variants are determined by coset decomposition with an intersection group that depends on the point groups of the phases and on the type of transformation matrix [20]. In the CT, the most important variants are those obtained by correspondence. For the reference variant for which the correspondence matrix $\mathbf{C}^{A \to M}$ has been determined, it can be shown that some symmetries of the parent austenite are preserved by correspondence, i.e. these symmetries become after lattice distortion symmetries of the daughter phase. They form the correspondence intersection subgroup between the point group of austenite $\mathbb{G}^{\mathrm{A}}$ and the point group of martensite $\mathbb{G}^{\mathrm{M}}$. This subgroup is given by

$$\mathbb{H}_{\mathrm{C}}^{\mathrm{A}} = \mathbb{G}^{\mathrm{A}} \cap \mathbf{C}^{\mathrm{A}\to\mathrm{M}}\, \mathbb{G}^{\mathrm{M}}\, \mathbf{C}^{\mathrm{M}\to\mathrm{A}} \tag{25}$$

In other words, $\mathbb{H}_{\mathrm{C}}^{\mathrm{A}}$ is constituted of the parent and daughter symmetries that are in correspondence each other. Note that this correspondence does not necessarily imply a parallelism of the symmetry elements. The correspondence variants $\mathrm{M}_i$ are defined by the left cosets $\mathbf{g}_i^{\mathrm{A}}\, \mathbb{H}_{\mathrm{C}}^{\mathrm{A}}$, with their associated set of equivalent correspondence matrices $\{\mathbf{C}^{\mathrm{A}\to\mathrm{M}_i}\}$ given by

$$\{\mathbf{C}^{\mathrm{A}\to\mathrm{M}_i}\} = \mathbf{g}_i^{\mathrm{A}}\, \mathbb{H}_{\mathrm{C}}^{\mathrm{A}}\, \mathbf{C}^{\mathrm{A}\to\mathrm{M}} \tag{26}$$

The matrices $\mathbf{g}_i^{\mathrm{A}}$ are the symmetry elements of the point group $\mathbb{G}^{\mathrm{A}}$. It is implicitly assumed that $\mathbf{g}_1^{\mathrm{A}}$ is the identity matrix in $\mathbb{G}^{\mathrm{A}}$. More details can be found in Ref. [23]. The number of correspondence variants of martensite is given by Lagrange's formula,

$$N_{\mathrm{C}}^{\mathrm{M}} = \frac{|\mathbb{G}^{\mathrm{A}}|}{|\mathbb{H}_{\mathrm{C}}^{\mathrm{A}}|} \tag{27}$$

It is often written in the literature that the number of variants is $\frac{|\mathbb{G}^{\mathrm{A}}|}{|\mathbb{G}^{\mathrm{M}}|}$, but this formula is vague and in general incorrect because the type of variants (correspondence, orientation, distortion, or stretch) is not specified and because the intersection group is not necessarily isomorph to the martensite point group. The shape memory effect is a direct consequence of the fact that the reverse transformation produces only one austenite correspondence variant. Let us prove it. Since $\mathbb{H}_{\mathrm{C}}^{\mathrm{M}} = \mathbb{G}^{\mathrm{M}} \cap \mathbf{C}^{\mathrm{M}\to\mathrm{A}}\, \mathbb{G}^{\mathrm{A}}\, \mathbf{C}^{\mathrm{A}\to\mathrm{M}} = \mathbf{C}^{\mathrm{M}\to\mathrm{A}}\, \mathbb{H}_{\mathrm{C}}^{\mathrm{A}}\, \mathbf{C}^{\mathrm{A}\to\mathrm{M}}$; an isomorphism exists between the two subgroups $\mathbb{H}_{\mathrm{C}}^{\mathrm{M}}$ and $\mathbb{H}_{\mathrm{C}}^{\mathrm{A}}$, thus $|\mathbb{H}_{\mathrm{C}}^{\mathrm{M}}| = |\mathbb{H}_{\mathrm{C}}^{\mathrm{A}}|$, i.e. the correspondence subgroups contain the same number of symmetries for direct and reverse transformations. In the case of shape memory alloys, all the symmetries of martensite are inherited by correspondence from the symmetries of austenite; i.e. $|\mathbb{G}^{\mathrm{M}}| = |\mathbb{H}_{\mathrm{C}}^{\mathrm{A}}|$. Applying formula (27) to the reverse transformation leads to $N_{\mathrm{C}}^{\mathrm{A}} = \frac{|\mathbb{G}^{\mathrm{M}}|}{|\mathbb{H}_{\mathrm{C}}^{\mathrm{M}}|}$; and since $|\mathbb{H}_{\mathrm{C}}^{\mathrm{M}}| = |\mathbb{H}_{\mathrm{C}}^{\mathrm{A}}| = |\mathbb{G}^{\mathrm{M}}|$, we get

$$N_{\mathrm{C}}^{\mathrm{A}} = 1 \tag{28}$$

This shows that only one possible austenite variant is created by heating from any of the martensite variants. If the accommodation between the martensite variants is fully elastic, the material necessarily comes back to the initial austenite orientation, which explains the reversibility of the transformation, and the shape memory effect.

## 3.5 The different types of intercorrespondences between the martensite variants

We have seen in section 3.3, how the geometrical elements of the type I and type II transformation twins ("shear" plane, direction amplitude) can be calculated from the correspondence matrix and the reflection and 180° rotation symmetry matrices of the austenite. There is no need to do the calculations for all the 2-fold symmetry elements because we have shown that the calculations are based on the intercorrespondence matrices, and it is possible to group them to avoid redundant calculations. We consider two correspondence variants $(\mathrm{M}_i, \mathrm{M}_j)$. The equivalent intercorrespondence matrices between them is given by the set of matrices

$$\{\mathbf{C}^{\mathrm{M}_i\to\mathrm{M}_j}\} = \{\mathbf{C}^{\mathrm{M}_i\to\mathrm{A}}\, \mathbf{C}^{\mathrm{A}\to\mathrm{M}_j}\} = \mathbf{C}^{\mathrm{M}\to\mathrm{A}}\, \mathbb{H}_{\mathrm{T}}^{\mathrm{A}}\, \mathbf{g}_k^{\mathrm{A}}\, \mathbb{H}_{\mathrm{C}}^{\mathrm{A}}\, \mathbf{C}^{\mathrm{A}\to\mathrm{M}} \tag{29}$$

where $\mathbf{g}_k^{\mathrm{A}} = (\mathbf{g}_i^{\mathrm{A}})^{-1}\mathbf{g}_j^{\mathrm{A}} \in \mathbb{G}^{\mathrm{A}}$.

We call "intercorrespondence operator" the double-coset $\mathbf{O}_k = \mathbb{H}_C^A\, \mathbf{g}_k^A\, \mathbb{H}_C^A$. The intercorrespondence matrices are formed from the $\mathbf{O}_k$ by the isomorphism $\mathbf{C}^{M\to A}\, \mathbf{O}_k\, \mathbf{C}^{A\to M}$. The double-cosets were introduced for the first time in crystallography by Janovec in his research on ferroelectric domains [24,25]. The number of different types of intercorrespondence is the number of double-cosets; it is given by Burnside's formula [23]. Once the parent symmetry matrices $\mathbf{g}_k^A$ in the intercorrespondence operators $\mathbf{O}_k$ are determined, the transformation twins can be calculated with the equations of section 3.3. If the operator contains a mirror symmetry on a parent plane $\mathbf{p}_A$, a type I twin can be established. If the operator contains a 2-fold rotation around a parent plane $\mathbf{a}_A$, a type II twin can be established. Note that partitioning the austenite point group $\mathbb{G}^A$ into the different correspondence left cosets and intercorrespondence double-cosets is quite simple because the symmetry matrices are written in the crystallographic basis and are thus constituted only of 1, 0 or/and -1. Step-by-step explanations are given in the Appendix A of Ref. [19]. Operators that contain at least one two-fold symmetry are called "ambivalent"; if not, they are called "polar". The twins in polar operators are necessarily "weak"; they are not considered in the present paper. An example of double-coset decomposition and twin classification is given with the B2 → B19′ transformation in NiTi shape memory alloy in Table 1. The operators $\mathbf{O}_2$, $\mathbf{O}_4$, $\mathbf{O}_4$, $\mathbf{O}_5$ and $\mathbf{O}_6$ are ambivalent. The operator $\mathbf{O}_2$ only contains compound twins. The operators $\mathbf{O}_1$ and $\mathbf{O}_3$ are complementary polar operators. More details can be found in Ref. [19].

Table 1. Intercorrespondence operators with their symmetries matrices in the case of the B2 → B19′ transformation in NiTi, from Ref. [19]. The 2-fold symmetries (reflections and 180° rotations) from which the generic elements of the type I and type II twins are deduced are marked in green.

| | | B2 symmetries in the double-cosets | | Twin element (non generic marked by ~) | shear |
|---|---|---|---|---|---|
| | **Disorient.** | **Matrices** | **Geometrical element** | | |
| $\mathbf{O}_0$ | I | $\begin{pmatrix}1&0&0\\0&1&0\\0&0&1\end{pmatrix}\begin{pmatrix}-1&0&0\\0&-1&0\\0&0&-1\end{pmatrix}\begin{pmatrix}0&-1&0\\-1&0&0\\0&0&1\end{pmatrix}\begin{pmatrix}0&1&0\\1&0&0\\0&0&-1\end{pmatrix}$ | I, $\bar{\mathrm{I}}$, $\boldsymbol{m}^{\mathbf{B2}}_{(\mathbf{110})}$, $\boldsymbol{R}^{\mathbf{B2}}_{\boldsymbol{\pi},[\mathbf{110}]}$ | | |
| $\mathbf{O}_2$ | $R^{B19'}_{\pi,[001]}$ | $\begin{pmatrix}1&0&0\\0&1&0\\0&0&-1\end{pmatrix}\begin{pmatrix}0&1&0\\1&0&0\\0&0&1\end{pmatrix}\begin{pmatrix}-1&0&0\\0&-1&0\\0&0&1\end{pmatrix}\begin{pmatrix}0&-1&0\\-1&0&0\\0&0&-1\end{pmatrix}$ | $\boldsymbol{m}^{\mathbf{B2}}_{(\mathbf{001})}$, $\boldsymbol{m}^{\mathbf{B2}}_{(\mathbf{1\bar{1}0})}$, $\boldsymbol{R}^{\mathbf{B2}}_{\boldsymbol{\pi},[\mathbf{001}]}$, $\boldsymbol{R}^{\mathbf{B2}}_{\boldsymbol{\pi},[\mathbf{1\bar{1}0}]}$ | Comp.1 : $(100)_{B19'} \parallel (001)_{B2}$<br>$[001]_{B19'} \parallel [\bar{1}10]_{B2}$<br>Comp.2 : $(001)_{B19'} \parallel (1\bar{1}0)_{B2}$<br>$[100]_{B19'} \parallel [001]_{B2}$ | $s = 0.2389$ |
| $\mathbf{O}_4$ | $R^{B19'}_{2\pi/3,\sim[17,0,16]}$ | $\begin{pmatrix}1&0&0\\0&0&1\\0&1&0\end{pmatrix}\begin{pmatrix}0&0&-1\\0&1&0\\-1&0&0\end{pmatrix}\begin{pmatrix}0&0&1\\1&0&0\\0&-1&0\end{pmatrix}\begin{pmatrix}0&1&0\\0&0&-1\\1&0&0\end{pmatrix}$<br>$\begin{pmatrix}-1&0&0\\0&0&-1\\0&-1&0\end{pmatrix}\begin{pmatrix}0&0&1\\0&-1&0\\1&0&0\end{pmatrix}\begin{pmatrix}0&0&-1\\-1&0&0\\0&1&0\end{pmatrix}\begin{pmatrix}0&-1&0\\0&0&1\\-1&0&0\end{pmatrix}$ | $\boldsymbol{m}^{\mathbf{B2}}_{(\mathbf{01\bar{1}})}$, $\boldsymbol{m}^{\mathbf{B2}}_{(\mathbf{101})}$, $\bar{\boldsymbol{R}}^{\mathrm{B2}}_{2\pi/3,[1\bar{1}\bar{1}]}$, $\bar{\boldsymbol{R}}^{\mathrm{B2}}_{-2\pi/3,[1\bar{1}\bar{1}]}$<br>$\boldsymbol{R}^{\mathbf{B2}}_{\boldsymbol{\pi},[\mathbf{01\bar{1}}]}$, $\boldsymbol{R}^{\mathbf{B2}}_{\boldsymbol{\pi},[\mathbf{101}]}$, $\boldsymbol{R}^{\mathrm{B2}}_{2\pi/3,[1\bar{1}\bar{1}]}$, $\boldsymbol{R}^{\mathrm{B2}}_{-2\pi/3,[1\bar{1}\bar{1}]}$ | type I: $(\bar{1}11)_{B19'} \parallel (01\bar{1})_{B2}$<br>$\sim[7,\bar{6},13]_{B19'} \parallel [\overline{19},7,7]_{B2}$<br>type II: $[\bar{2}11]_{B19'} \parallel [01\bar{1}]_{B2}$<br>$\sim(1\ \bar{2}\ 4)_{B19'} \parallel (\bar{3}\ 1\ 1)_{B2}$ | $s = 0.3096$ |
| $\mathbf{O}_5$ | $R^{B19'}_{2\pi/3,\sim[40\bar{3}]}$ | $\begin{pmatrix}1&0&0\\0&0&-1\\0&-1&0\end{pmatrix}\begin{pmatrix}0&0&1\\0&1&0\\1&0&0\end{pmatrix}\begin{pmatrix}0&0&-1\\1&0&0\\0&1&0\end{pmatrix}\begin{pmatrix}0&1&0\\0&0&1\\-1&0&0\end{pmatrix}$<br>$\begin{pmatrix}-1&0&0\\0&0&1\\0&1&0\end{pmatrix}\begin{pmatrix}0&0&-1\\0&-1&0\\-1&0&0\end{pmatrix}\begin{pmatrix}0&-1&0\\0&0&-1\\1&0&0\end{pmatrix}\begin{pmatrix}0&0&1\\-1&0&0\\0&-1&0\end{pmatrix}$ | $\boldsymbol{m}^{\mathbf{B2}}_{(\mathbf{011})}$, $\boldsymbol{m}^{\mathbf{B2}}_{(\mathbf{\bar{1}01})}$, $\bar{\boldsymbol{R}}^{\mathrm{B2}}_{2\pi/3,[\bar{1}1\bar{1}]}$, $\bar{\boldsymbol{R}}^{\mathrm{B2}}_{-2\pi/3,[\bar{1}1\bar{1}]}$<br>$\boldsymbol{R}^{\mathbf{B2}}_{\boldsymbol{\pi},[\mathbf{011}]}$, $\boldsymbol{R}^{\mathbf{B2}}_{\boldsymbol{\pi},[\mathbf{\bar{1}01}]}$, $\boldsymbol{R}^{\mathrm{B2}}_{-2\pi/3,[\bar{1}1\bar{1}]}$, $\boldsymbol{R}^{\mathrm{B2}}_{2\pi/3,[\bar{1}1\bar{1}]}$ | type I: $(111)_{B19'} \parallel (011)_{B2}$<br>$\sim[\bar{6}\ 2\ 4]_{B19'} \parallel [\bar{1},3,\bar{3}]_{B2}$<br>type II: $[211]_{B19'} \parallel [011]_{B2}$<br>$\sim(\bar{2}\ 1\ 3)_{B19'} \parallel (\bar{1}\ 2\ \bar{2})_{B2}$ | $s = 0.1422$ |
| $\mathbf{O}_6$ | $R^{B19'}_{\pi/2,\sim[11,0,1]}$ | $\begin{pmatrix}1&0&0\\0&-1&0\\0&0&1\end{pmatrix}\begin{pmatrix}-1&0&0\\0&1&0\\0&0&1\end{pmatrix}\begin{pmatrix}0&1&0\\-1&0&0\\0&0&-1\end{pmatrix}\begin{pmatrix}0&-1&0\\1&0&0\\0&0&-1\end{pmatrix}$<br>$\begin{pmatrix}-1&0&0\\0&1&0\\0&0&-1\end{pmatrix}\begin{pmatrix}1&0&0\\0&-1&0\\0&0&-1\end{pmatrix}\begin{pmatrix}0&-1&0\\1&0&0\\0&0&1\end{pmatrix}\begin{pmatrix}0&1&0\\-1&0&0\\0&0&1\end{pmatrix}$ | $\boldsymbol{m}^{\mathbf{B2}}_{(\mathbf{010})}$, $\boldsymbol{m}^{\mathbf{B2}}_{(\mathbf{100})}$, $\bar{\boldsymbol{R}}^{\mathrm{B2}}_{\pi/2,[001]}$, $\bar{\boldsymbol{R}}^{\mathrm{B2}}_{-\pi/2,[001]}$<br>$\boldsymbol{R}^{\mathbf{B2}}_{\boldsymbol{\pi},[\mathbf{010}]}$, $\boldsymbol{R}^{\mathbf{B2}}_{\boldsymbol{\pi},[\mathbf{100}]}$, $\boldsymbol{R}^{\mathrm{B2}}_{\pi/2,[001]}$, $\boldsymbol{R}^{\mathrm{B2}}_{-\pi/2,[001]}$ | type I: $(011)_{B19'} \parallel (010)_{B2}$<br>$\sim[\overline{11},\bar{7},7]_{B19'}$<br>$\parallel [\overline{14},0,\overline{11}]_{B2}$<br>type II: $[011]_{B19'} \parallel [010]_{B2}$<br>$\sim(3\ 4\ \bar{4})_{B19'} \parallel (4\ 0\ 3)_{B2}$ | $s = 0.2804$ |
| $\mathbf{O}_1$ | $R^{B19'}_{-\pi/2,\sim[0,8,7]}$ | $\begin{pmatrix}1&0&0\\0&0&1\\0&-1&0\end{pmatrix}\begin{pmatrix}0&0&-1\\0&1&0\\1&0&0\end{pmatrix}\begin{pmatrix}0&0&1\\1&0&0\\0&1&0\end{pmatrix}\begin{pmatrix}0&1&0\\0&0&-1\\-1&0&0\end{pmatrix}$<br>$\begin{pmatrix}-1&0&0\\0&0&-1\\0&1&0\end{pmatrix}\begin{pmatrix}0&0&1\\0&-1&0\\-1&0&0\end{pmatrix}\begin{pmatrix}0&0&-1\\-1&0&0\\0&-1&0\end{pmatrix}\begin{pmatrix}0&-1&0\\0&0&1\\1&0&0\end{pmatrix}$ | $\boldsymbol{R}^{\mathrm{B2}}_{-\pi/2,[100]}$, $\boldsymbol{R}^{\mathrm{B2}}_{-\pi/2,[010]}$, $\boldsymbol{R}^{\mathrm{B2}}_{2\pi/3,[111]}$, $\boldsymbol{R}^{\mathrm{B2}}_{-2\pi/3,[\bar{1}\bar{1}1]}$<br>$\bar{\boldsymbol{R}}^{\mathrm{B2}}_{-\pi/2,[100]}$, $\bar{\boldsymbol{R}}^{\mathrm{B2}}_{-\pi/2,[010]}$, $\bar{\boldsymbol{R}}^{\mathrm{B2}}_{2\pi/3,[111]}$, $\bar{R}^{\mathrm{B2}}_{-2\pi/3,[\bar{1}\bar{1}1]}$ | weak 1: $[011]_{B19'} \parallel [010]_{B2}$ | $s_g = 0.2911$ |
| $\mathbf{O}_3$ | $R^{B19'}_{\pi/2,\sim[0,8,7]}$ | $\begin{pmatrix}1&0&0\\0&0&-1\\0&1&0\end{pmatrix}\begin{pmatrix}0&0&1\\0&1&0\\-1&0&0\end{pmatrix}\begin{pmatrix}0&1&0\\0&0&1\\1&0&0\end{pmatrix}\begin{pmatrix}0&0&-1\\1&0&0\\0&-1&0\end{pmatrix}$<br>$\begin{pmatrix}-1&0&0\\0&0&1\\0&-1&0\end{pmatrix}\begin{pmatrix}0&0&-1\\0&-1&0\\1&0&0\end{pmatrix}\begin{pmatrix}0&-1&0\\0&0&-1\\-1&0&0\end{pmatrix}\begin{pmatrix}0&0&1\\-1&0&0\\0&1&0\end{pmatrix}$ | $\boldsymbol{R}^{\mathrm{B2}}_{\pi/2,[100]}$, $\boldsymbol{R}^{\mathrm{B2}}_{\pi/2,[010]}$, $R^{\mathrm{B2}}_{-2\pi/3,[111]}$, $R^{\mathrm{B2}}_{2\pi/3,[\bar{1}\bar{1}1]}$<br>$\bar{\boldsymbol{R}}^{\mathrm{B2}}_{\pi/2,[100]}$, $\bar{\boldsymbol{R}}^{\mathrm{B2}}_{\pi/2,[010]}$, $\bar{\boldsymbol{R}}^{\mathrm{B2}}_{-2\pi/3,[111]}$, $\bar{\boldsymbol{R}}^{\mathrm{B2}}_{2\pi/3,[\bar{1}\bar{1}1]}$ | weak 2: $[011]_{B19'} \parallel [010]_{B2}$ | |

It was checked that all the components of type I and type II twins (directions and planes) calculated by the equations described in section 3.3 are numerically equal to those calculated from the PTMC.

Now, we will show that the correspondence, metrics and group of symmetries that have been used to calculate the transformation twins can be used to establish the specific lattice parameters that verify the supercompatibility conditions.

# 4 Supercompatibility conditions from correspondence

## 4.1 A/M compatibility written as a degeneracy condition

We have shown in section 2 that the three supercompatibility conditions (SC 1,2,3) can be replaced by three conditions: A/M compatibility (IPS lattice distortion), M/M compatibility (twin relation), and a shear/shear relation. These three conditions allow for a perfect A/M/M compatibility. Let us express in this subsection the A/M compatibility condition from the correspondence matrix. We recall from section 3 that any austenite direction $\mathbf{u}_{\mathrm{A}}$ becomes after lattice distortion a martensite direction $\mathbf{u}_{\mathrm{M}} = \mathbf{C}^{\mathrm{M}\to\mathrm{A}}\,\mathbf{u}_{\mathrm{A}}$. The square of the norm of $\mathbf{u}_{\mathrm{A}}$ is $\|\mathbf{u}_{\mathrm{A}}\|^2 = \mathbf{u}_{\mathrm{A}}^{\mathrm{t}}\,\boldsymbol{\mathcal{M}}_{\mathrm{A}}\,\mathbf{u}_{\mathrm{A}}$. The square of the norm of $\mathbf{u}_{\mathrm{M}}$ is $\|\mathbf{u}_{\mathrm{M}}\|^2 = \mathbf{u}_{\mathrm{M}}^{\mathrm{t}}\,\boldsymbol{\mathcal{M}}_{\mathrm{M}}\,\mathbf{u}_{\mathrm{M}} = \mathbf{u}_{\mathrm{A}}^{\mathrm{t}}\left(\left(\mathbf{C}^{\mathrm{M}\to\mathrm{A}}\right)^{\mathrm{t}}\boldsymbol{\mathcal{M}}_{\mathrm{M}}\,\mathbf{C}^{\mathrm{M}\to\mathrm{A}}\right)\mathbf{u}_{\mathrm{A}}$. The norm of $\mathbf{u}_{\mathrm{A}}$ does not change by phase transformation if and only if $\|\mathbf{u}_{\mathrm{M}}\|^2 = \|\mathbf{u}_{\mathrm{A}}\|^2$, i.e. $\mathbf{u}_{\mathrm{M}}^{\mathrm{t}}\,\boldsymbol{\mathcal{M}}_{\mathrm{M}}\,\mathbf{u}_{\mathrm{M}} = \mathbf{u}_{\mathrm{A}}^{\mathrm{t}}\,\boldsymbol{\mathcal{M}}_{\mathrm{A}}\,\mathbf{u}_{\mathrm{A}}$. The invariance of the norm of $\mathbf{u}_{\mathrm{A}}$ is thus given by

$$\mathbf{u}_{\mathrm{A}}^{\mathrm{t}}\,\mathbf{CMC}\,\mathbf{u}_{\mathrm{A}} = 0 \tag{30}$$

$$\text{with } \mathbf{CMC} = \left(\mathbf{C}^{\mathrm{M}\to\mathrm{A}}\right)^{\mathrm{t}}\boldsymbol{\mathcal{M}}_{\mathrm{M}}\,\mathbf{C}^{\mathrm{M}\to\mathrm{A}} - \boldsymbol{\mathcal{M}}_{\mathrm{A}} \tag{31}$$

where CMC stands for "compatibility by metric correspondence". Noting $[x, y, z]$ the coordinates of $\mathbf{u}_{\mathrm{A}}$, equation (30) takes the general quadratic form

$$q_{\mathrm{CMC}}(x, y, z) = q_{11}\,x^2 + q_{22}\,y^2 + q_{33}\,z^2 + 2q_{12}\,x\,y + 2q_{23}\,y\,z + 2q_{13}\,x\,z = 0 \tag{32}$$

with $q_{ij} = \mathbf{CMC}_{(i,j)}$. The solutions of the equation form a surface $\mathcal{S}_{\mathrm{CMC}}$ constituted by all the vectors $\mathbf{u}_{\mathrm{A}} = [x, y, z]$ of norm preserved by correspondence. Equation (32) has real solutions only if some coefficients $q_{ij}$ of the polynomial form have opposite signs. It is a specific hyperboloid constituted by rays that all cross the origin. Indeed, if $[x, y, z]$ is a solution, then $r[x, y, z]$ is also a solution, whatever the $r \in \mathbb{R}$. Thus, $\mathcal{S}_{\mathrm{CMC}}$ is a double-cone. Its symmetries form a subgroup of the austenite symmetries $\mathbf{g}^{\mathrm{A}} \in \mathbb{G}^{\mathrm{A}}$ such that $\forall\, \mathbf{u}_{\mathrm{A}} \in \mathcal{S}_{CMC} \Rightarrow \mathbf{g}^{\mathrm{A}}\,\mathbf{u}_{\mathrm{A}} \in \mathcal{S}_{CMC}$. It can be defined by

$$\mathbb{G}_{\mathrm{CMC}} = \left\{\mathbf{g}^{\mathrm{A}} \in \mathbb{G}^{\mathrm{A}}, \mathbf{u}_{\mathrm{A}}^{\mathrm{t}}\left(\mathbf{g}^{\mathrm{A}}\right)^{\mathrm{t}}\mathbf{CMC}\;\mathbf{g}^{\mathrm{A}}\,\mathbf{u}_{\mathrm{A}} = 0, \forall\, \mathbf{u}_{\mathrm{A}} \in \mathcal{S}_{CMC}\right\} \tag{33}$$

$\mathbb{G}_{\mathrm{CMC}}$ is a subgroup of $\mathbb{G}^{\mathrm{A}}$ that contains the correspondence subgroup $\mathbb{H}_{\mathrm{C}}^{\mathrm{A}}$ given by equation (25). To show it, we notice first that, by equation (16), the metric tensor is stable by symmetry, $\forall\, \mathbf{g}^{\mathrm{A}} \in \mathbb{G}^{\mathrm{A}}$, $\left(\mathbf{g}^{\mathrm{A}}\right)^{\mathrm{t}}\boldsymbol{\mathcal{M}}_{\mathrm{A}}\,\mathbf{g}^{\mathrm{A}} = \boldsymbol{\mathcal{M}}_{\mathrm{A}}$, and $\forall\, \mathbf{g}^{\mathrm{M}} \in \mathbb{G}^{\mathrm{M}}$, $\left(\mathbf{g}^{\mathrm{M}}\right)^{\mathrm{t}}\boldsymbol{\mathcal{M}}_{\mathrm{M}}\,\mathbf{g}^{\mathrm{M}} = \boldsymbol{\mathcal{M}}_{\mathrm{M}}$. Then, if $\mathbf{g}^{\mathrm{A}} \in \mathbb{H}_{\mathrm{C}}^{\mathrm{A}}$, $\exists\, \mathbf{g}^{\mathrm{M}} \in \mathbb{G}^{\mathrm{M}}$, $\mathbf{g}^{\mathrm{A}} = \mathbf{C}^{\mathrm{A}\to\mathrm{M}}\,\mathbf{g}^{\mathrm{M}}\,\mathbf{C}^{\mathrm{M}\to\mathrm{A}}$; thus, $\forall\, \mathbf{u}_{\mathrm{A}} \in \mathcal{S}_{CMC}, \mathbf{u}_{\mathrm{A}}^{\mathrm{t}}\left(\mathbf{g}^{\mathrm{A}}\right)^{\mathrm{t}}\mathbf{CMC}\,\mathbf{g}^{\mathrm{A}}\,\mathbf{u}_{\mathrm{A}} = \mathbf{u}_{\mathrm{A}}^{\mathrm{t}}\left(\mathbf{C}^{\mathrm{M}\to\mathrm{A}}\right)^{\mathrm{t}}\left(\mathbf{g}^{\mathrm{M}}\right)^{\mathrm{t}}\boldsymbol{\mathcal{M}}_{\mathrm{M}}\,\mathbf{g}^{\mathrm{M}}\,\mathbf{C}^{\mathrm{M}\to\mathrm{A}}\,\mathbf{u}_{\mathrm{A}} - \mathbf{u}_{\mathrm{A}}^{\mathrm{t}}\,\boldsymbol{\mathcal{M}}_{\mathrm{A}}\,\mathbf{u}_{\mathrm{A}} = \mathbf{u}_{\mathrm{A}}^{\mathrm{t}}\left(\mathbf{C}^{\mathrm{M}\to\mathrm{A}}\right)^{\mathrm{t}}\boldsymbol{\mathcal{M}}_{\mathrm{M}}\,\mathbf{C}^{\mathrm{M}\to\mathrm{A}}\;\mathbf{u}_{\mathrm{A}} - \mathbf{u}_{\mathrm{A}}^{\mathrm{t}}\,\boldsymbol{\mathcal{M}}_{\mathrm{A}}\,\mathbf{u}_{\mathrm{A}} = 0 \Rightarrow \mathbf{g}^{\mathrm{A}} \in \mathbb{G}_{\mathrm{CMC}}$.

Actually, in general $\mathbb{G}_{CMC} = \mathbb{H}_{\mathrm{C}}^{\mathrm{A}}$, but for some specific metrics, by "accident" or by "design", it is possible that $\mathbb{H}_{\mathrm{C}}^{\mathrm{A}} < \mathbb{G}_{\mathrm{CMC}}$, which means that $\mathbb{H}_{\mathrm{C}}^{\mathrm{A}}$ is a subgroup of $\mathbb{G}_{\mathrm{CMC}}$, but $\mathbb{H}_{\mathrm{C}}^{\mathrm{A}} \neq \mathbb{G}_{\mathrm{CMC}}$. We will show that this happens when the quadratic form is degenerated, and austenite/martensite compatibility is reached.

Since the **CMC** matrix is symmetric, it can be diagonalized inside an orthonormal eigen basis $\boldsymbol{\mathcal{B}}_d = (\boldsymbol{eg}_1, \boldsymbol{eg}_2, \boldsymbol{eg}_3)$ constituted of the **CMC** eigenvectors. The eigenvalues $(q_1, q_2, q_3)$ are the roots of the characteristic polynomial equation $\det(\mathbf{CMC} - q\,\mathbf{I}) = 0$. We consider a vector $\mathbf{u}$ of coordinates $[x, y, z]$ in the basis $\boldsymbol{\mathcal{B}}_c^{\mathrm{A}}$ and $[X, Y, Z]$ in the orthonormal basis $\boldsymbol{\mathcal{B}}_d$. The coordinate transformation matrix $\mathbf{P} =$ $[\boldsymbol{\mathcal{B}}_c^{\mathrm{A}} \to \boldsymbol{\mathcal{B}}_d]$ is obtained by writing the three vectors $(\boldsymbol{eg}_1, \boldsymbol{eg}_2, \boldsymbol{eg}_3)$ in columns. These coordinates of are linked by $\begin{bmatrix} x \\ y \\ z \end{bmatrix} = \mathbf{P} \begin{bmatrix} X \\ Y \\ Z \end{bmatrix}$. In the basis $\boldsymbol{\mathcal{B}}_d$, equation (32) becomes

$$[X, Y, Z] \; diag\,(q_1, q_2, q_3) \begin{bmatrix} X \\ Y \\ Z \end{bmatrix} = q_1 X^2 + q_2 Y^2 + q_3 Z^2 = 0 \qquad (34)$$

The solution to this quadratic equation is not reduced to (0,0,0) if and only if one of the three values $(q_1, q_2, q_3)$ has a sign opposite to the two others. A/M compatibility is obtained when all the vectors $\mathbf{u}_{\mathrm{A}}$ in the habit plane have their norm invariant, i.e. their coordinates $[x, y, z]$ should verify equation (32), or by a change of basis $[X, Y, Z]$ should verify equation (34). This is possible when the double-cone of equation (34) is degenerated into a double-plane, as illustrated in Figure 5. This degeneracy condition can be written

$$q_i = 0 \;\;\&\;\; q_j\, q_k \leq 0\,, \text{with } (i, j, k) \in \{(1,2,3)\} \qquad (35)$$

where $\{(1,2,3)\}$ means the 6 sets equivalent to (1,2,3) by permutations. For example, if $q_2 = 0$, $q_1 \geq 0$ and $q'_3 = -q_3 \geq 0$, the CMC equation is $q_1 X^2 - q'_3 Z^2 = \left(\sqrt{q_1} X + \sqrt{q'_3}\, Z\right)\left(\sqrt{q_1} X - \sqrt{q'_3}\, Z\right) = 0$, which show that the two possible habit planes indexed in $\boldsymbol{\mathcal{B}}_d$ are $\left(\sqrt{q_1}, 0, \sqrt{q'_3}\right)$ and $\left(\sqrt{q_1}, 0, -\sqrt{q'_3}\right)$.

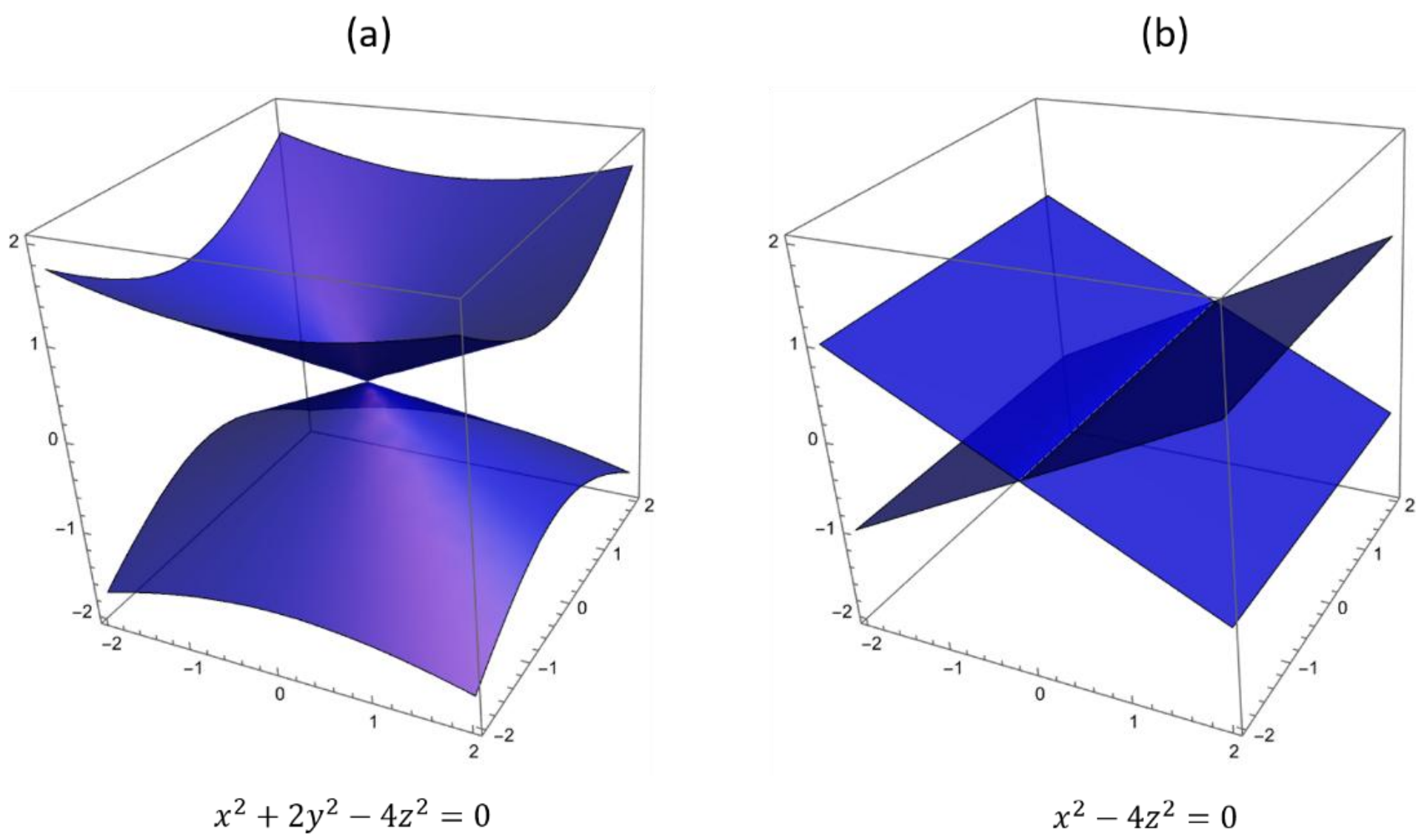

Figure 5. Example of CMC double-cone degeneracy. (a) Double-cone obtained when no specific condition is imposed to the lattice distortion. (b) Degeneracy of the double-cone into double-plane when the lattice distortion is an IPS.

We say that degeneracy is of *first order* when the inequality in the equations (35) is strict, i.e. $q_j q_k < 0$, and is of *second order* when $q_i = 0$ and ($q_j = 0$ or $q_k = 0$), which means that two eigenvalues of the **CMC** matrix are null. The degeneracy of second order corresponds to the case where the double-plane is reduced to a unique plane. This plane is necessarily given by the equation $X = 0$ for $q_2 = q_3 = 0$, $Y = 0$ for $q_1 = q_3 = 0$, or $Z = 0$ for $q_1 = q_2 = 0$. In other words:

$$q_i = q_j = 0, \text{with } (i, j) \in \{1,2,3\}^2 \;\&\; i \neq j \qquad (36)$$

When the degeneracy conditions of order 1 or 2 is obtained, the CMC equation (32) is reduced to a double or simple linear equation, respectively:

$q_{\mathrm{CMC}}(x,y,z) = (h_1\,x + k_1\,y + l_1\,z)(h_2\,x + k_2\,y + l_2\,z) = 0$ (order 1) (37)

$q_{\mathrm{CMC}}(x,y,z) = (h\,x + k\,y + l\,z)^2 = 0$ (order 2)

For order 1 degeneracy, two habit planes are possible: $\mathbf{m}_{\mathrm{A}} = (h_1\,k_1\,l_1)$ or $\mathbf{m}_{\mathrm{A}} = (h_2\,k_2\,l_2)$. For order 2 degenercy, only one habit plane is possible $\mathbf{m}_{\mathrm{A}} = (h\,k\,l)$.

The degeneracy of *third order* corresponds to the case

$$q_1 = q_2 = q_3 = 0 \quad (38)$$

For such level of degeneracy, the CMC matrix is null, which means that the metrics of the martensite perfectly matches by correspondence with the metrics of the austenite; all the vectors $(x,y,z)$ of $\mathbb{R}^3$ verify equation (30). The condition of degeneracy of third order can also be found by applying a double derivative to the quadratic form (32).

$$\frac{\partial q_{\mathrm{CMC}}(x,y,z)}{\partial x^2} = \frac{\partial q_{\mathrm{CMC}}(x,y,z)}{\partial y^2} = \frac{\partial q_{\mathrm{CMC}}(x,y,z)}{\partial z^2} = \frac{\partial q_{\mathrm{CMC}}(x,y,z)}{\partial xy} = \frac{\partial q_{\mathrm{CMC}}(x,y,z)}{\partial xz} = 0$$

We have shown in this section that the A/M compatibility (IPS condition) is equivalent to the degeneracy of the CMC double-cone into a double-plane. There is a maximum of 6 possible degeneracy conditions; each of them is constituted by one equality and one inequality by equations (35). These equations directly involve the parent and daughter metrics, which makes their relative role in the supercompatibility conditions easier to understand than with SC equations (1)-(3).

## 4.2 Making the A/M and M/M compatibilities compatible

The M/M compatibility is obtained by the type I or type II transformation twins. Its CT equations are given in section 3.3. For type I twins the (rational) mirror plane of the twin is $\mathbf{p}_{\mathrm{M}}$ directly deduced from the austenite mirror plane $\mathbf{p}_{\mathrm{A}}$ and the correspondence matrix by equation (17). For type II twins, the (irrational) mirror plane is $\mathbf{jp}_{\mathrm{M}}$ deduced from the (rational) austenite 180° rotation axis $\mathbf{a}_{\mathrm{A}}$, the correspondence matrix and the martensite metric tensor by equation (23). We have seen in section 2, that the A/M and M/M compatibilities can coexist only if $\mathbf{d}_{\mathrm{A}}$, the shear direction of the IPS, and $\mathbf{a}$, the "shear" direction of the twin, are linked by the shear/shear equation (12). If this equation is verified, the A/M and M/M compatibilities are compatible together and supercompatiblity, i.e. full A/M/M compatibility, is reached. However, the vector $\mathbf{d}_{\mathrm{A}}$ is not yet known. We assume that the habit plane $\mathbf{m}_{\mathrm{A}}$ corresponding to the A/M compatibility has been determined from the degeneracy of the CMC quadratic form. It is the invariant plane of the IPS lattice distortion $\mathbf{F}_{\mathrm{A}}$. The vector $\mathbf{d}_{\mathrm{A}}$ can be calculated directly from $\mathbf{m}_{\mathrm{A}}$ such that $\mathbf{F}_{\mathrm{A}} = \mathbf{I} + \mathbf{d}_{\mathrm{A}}\mathbf{m}_{\mathrm{A}}^{\mathrm{t}}$, without explicitly calculating $\mathbf{F}_{\mathrm{A}}$. Indeed, since the plane $\mathbf{m}_{\mathrm{A}}$ is invariant, it becomes after transformation in the martensite basis the plane $\mathbf{m}_{\mathrm{M}} = (\mathbf{C}^{\mathrm{M}\to\mathrm{A}})^*\,\mathbf{m}_{\mathrm{A}} = (\mathbf{C}^{\mathrm{M}\to\mathrm{A}})^{-\mathrm{t}}\,\mathbf{m}_{\mathrm{A}}$ that remains parallel to $\mathbf{m}_{\mathrm{A}}$. The unit normal to the habit plane before transformation is $\mathbf{n}_{\mathrm{A}} = \boldsymbol{\mathcal{M}}_{\mathrm{A}}^{-1}\;\mathbf{m}_{\mathrm{A}}$ in Miller indices of austenite, and $\mathbf{n}_{\mathrm{M}} = \boldsymbol{\mathcal{M}}_{\mathrm{M}}^{-1}\;\mathbf{m}_{\mathrm{M}}$ in Miller indices of martensite. After transformation this vector becomes $\mathbf{n'}_{\mathrm{M}} = \mathbf{C}^{\mathrm{M}\to\mathrm{A}}\,\mathbf{n}_{\mathrm{A}}$. The shear direction of the IPS in martensite coordinates is $\mathbf{d}_{\mathrm{M}} = \mathbf{n'}_{\mathrm{M}} - \mathbf{n}_{\mathrm{M}} = \mathbf{C}^{\mathrm{M}\to\mathrm{A}}\,\boldsymbol{\mathcal{M}}_{\mathrm{A}}^{-1}\;\mathbf{m}_{\mathrm{A}} - \boldsymbol{\mathcal{M}}_{\mathrm{M}}^{-1}\,\left(\mathbf{C}^{\mathrm{M}\to\mathrm{A}}\right)^{-\mathrm{t}}\,\mathbf{m}_{\mathrm{A}} = \left(\mathbf{C}^{\mathrm{M}\to\mathrm{A}}\,\boldsymbol{\mathcal{M}}_{\mathrm{A}}^{-1} - \boldsymbol{\mathcal{M}}_{\mathrm{M}}^{-1}\,(\mathbf{C}^{\mathrm{M}\to\mathrm{A}})^{-\mathrm{t}}\right)\mathbf{m}_{\mathrm{A}}$. This direction is inherited from the austenite direction $\mathbf{d}_{\mathrm{A}} = \mathbf{C}^{\mathrm{A}\to\mathrm{M}}\,\mathbf{d}_{\mathrm{M}}$. This leads to

$$\mathbf{d}_{\mathrm{A}} = \mathbf{SMC}\,\mathbf{m}_{\mathrm{A}} \quad (39)$$

$$\text{with } \mathbf{SMC} = \boldsymbol{\mathcal{M}}_{\mathrm{A}}^{-1} - \mathbf{C}^{\mathrm{A}\to\mathrm{M}}\,\boldsymbol{\mathcal{M}}_{\mathrm{M}}^{-1}\,(\mathbf{C}^{\mathrm{A}\to\mathrm{M}})^{\mathrm{t}} \quad (40)$$

where SMC stands for "shear by metric correspondence". Note the SMC matrix is not the inverse of the CMC matrix given in equation (32).

Once $\mathbf{d_A}$ is calculated by equation (39) , it can be checked whether or not the shear/shear equation (12) is verified. This equation is recalled here with the notations used in the section:

$$2\,(\mathbf{m}_{\mathrm{A}}^{\mathrm{t}}\mathbf{n})\,\mathbf{d_A} = \mathbf{a} \tag{41}$$

where $(\mathbf{n}, \mathbf{a})$ is the twinning system (normal to the shear plane, and shear direction) linking the two martensite variants but written in the austenite basis. They are directly calculated from the transformation twin elements calculated in section 3.3. For type I twins on the plane $\mathbf{p}_{\mathrm{A}}$, $\mathbf{n} = \widetilde{\mathcal{M}_{\mathrm{A}}^{-1}\,\mathbf{p}_{\mathrm{A}}}$ and $\mathbf{a} = s\,\tilde{\mathbf{a}}_{\mathrm{A}}$. For type II twins around the axis $\mathbf{a}_{\mathrm{A}}$, $\mathbf{n} = \widetilde{\mathcal{M}_{\mathrm{A}}^{-1}\,\mathbf{j}\mathbf{p}_{\mathrm{A}}}$ and $\mathbf{a} = s\,\tilde{\mathbf{a}}_{\mathrm{A}}$. In equation (41), $(\mathbf{n}, \mathbf{a})$ have the same meaning as in the SC equations (2) and (3). This equation links the shear plane $\mathbf{m}_{\mathrm{A}}$ obtained from the A/M compatibility condition with the twin plane and direction required for the M/M compatibility; it makes the A/M and M/M compatibilities compatible, and is key to reach supercompatibility. The flowchart explaining the full process to calculate the supercompatibility conditions is given in Figure 6.

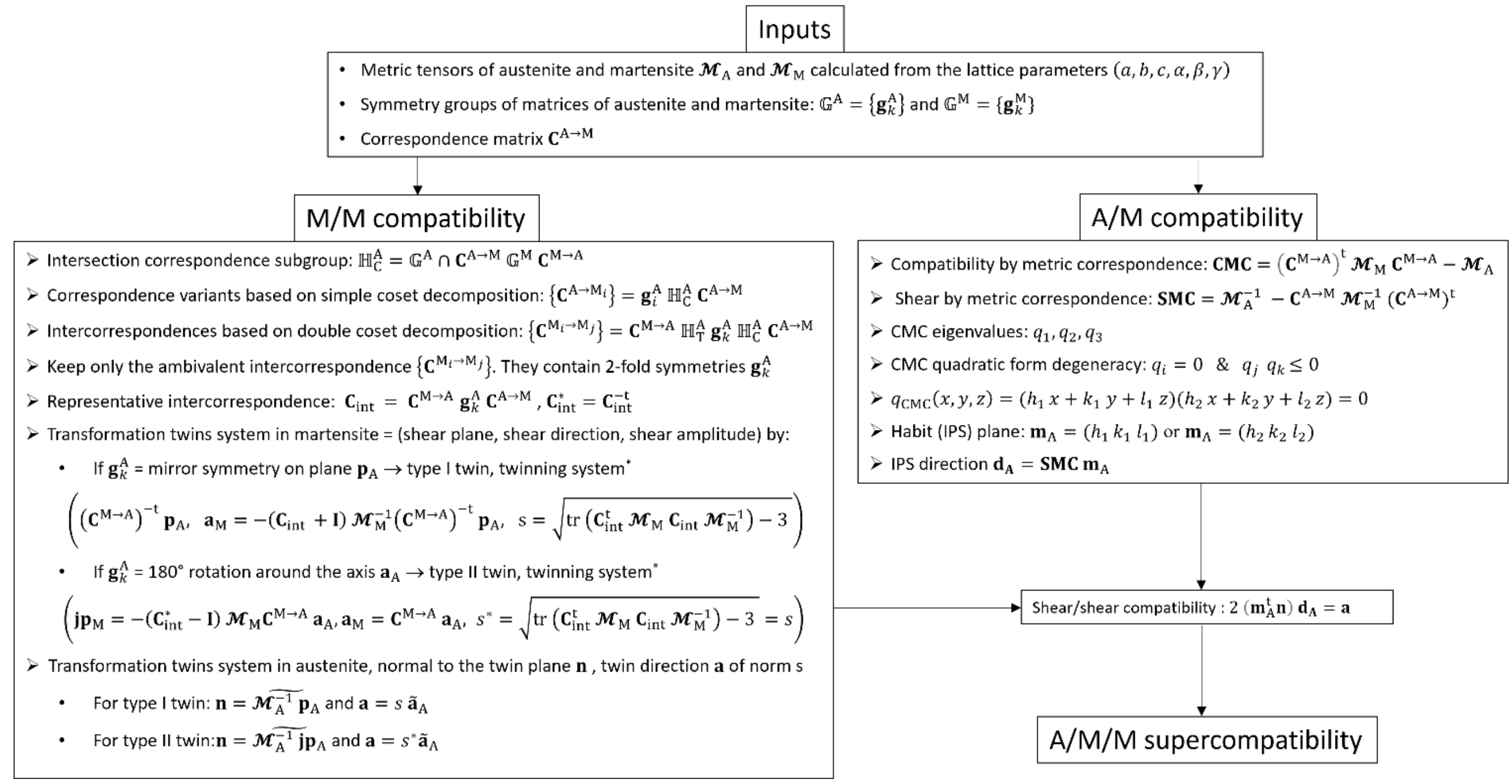

Figure 6. Flowchart of the CT calculations to determine the supercompatibility condutions, including the M/M transformation twins, the A/M coherent habit planes, and the shear/shear compatibility.

# 5 Supercompatibility conditions that could be targeted for the B2-B19' martensite transformation in NiTi alloys

## 5.1 Lattice parameters required for A/M compatibility

For martensitic transformation in NiTi, austenite A = B2 and martensite M = B19'. We consider $a_{B2}$ the lattice parameter of the cubic B2 phase, and $(a_{B19'}, b_{B19'}, c_{B19'}, \beta)$ the lattice parameters of the monoclinic B19' phase. The metrics of austenite is simply $\mathcal{M}_{\mathrm{A}} = a_{B2}^2\,\mathbf{I}$.

The metrics of martensite is $\boldsymbol{\mathcal{M}}_{\mathrm{M}} = \begin{pmatrix} a_{B19'}^2 & 0 & a_{B19'}\, c_{B19'} \cos(\beta) \\ 0 & b_{B19'}^2 & 0 \\ a_{B19'}\, c_{B19'}\, \cos(\beta) & 0 & c_{B19'}^2 \end{pmatrix}$.

The correspondence matrix following Otsuka and Ren's model [10] is $\mathbf{C}^{\mathrm{A}\to\mathrm{M}} = \begin{pmatrix} 0 & 1 & -1 \\ 0 & 1 & 1 \\ 1 & 0 & 0 \end{pmatrix}$, and its inverse $\mathbf{C}^{\mathrm{M}\to\mathrm{A}} = \begin{pmatrix} 0 & 0 & 1 \\ \frac{1}{2} & \frac{1}{2} & 0 \\ -\frac{1}{2} & \frac{1}{2} & 0 \end{pmatrix}$. To simplify the notations, we note $a = \frac{a_{B19'}}{a_{B2}}$, $b = \frac{b_{B19'}}{a_{B2}}$, $c = \frac{c_{B19'}}{a_{B2}}$.

The CMC matrix calculated from equation (31) is

$$\mathbf{CMC} = \begin{pmatrix} \frac{b^2+c^2}{4} - 1 & \frac{b^2-c^2}{4} & -\frac{1}{2} a\, c \cos(\beta) \\ \frac{b^2-c^2}{4} & \frac{b^2+c^2}{4} - 1 & \frac{1}{2} a\, c \cos(\beta) \\ -\frac{1}{2} a\, c \cos(\beta) & \frac{1}{2} a\, c \cos(\beta) & a^2 - 1 \end{pmatrix} \tag{42}$$

The equation (35) of its double-cone surface takes the form

$$c^2(x-y)^2 + b^2(x+y)^2 + 4a^2z^2 - 4\, a\, c\, (x-y)\, z \cos(\beta) - 4(x^2+y^2+z^2) = 0 \tag{43}$$

The surface obtained with the lattice parameters given by Kudoh *et al.* [26] is represented in Figure 7.

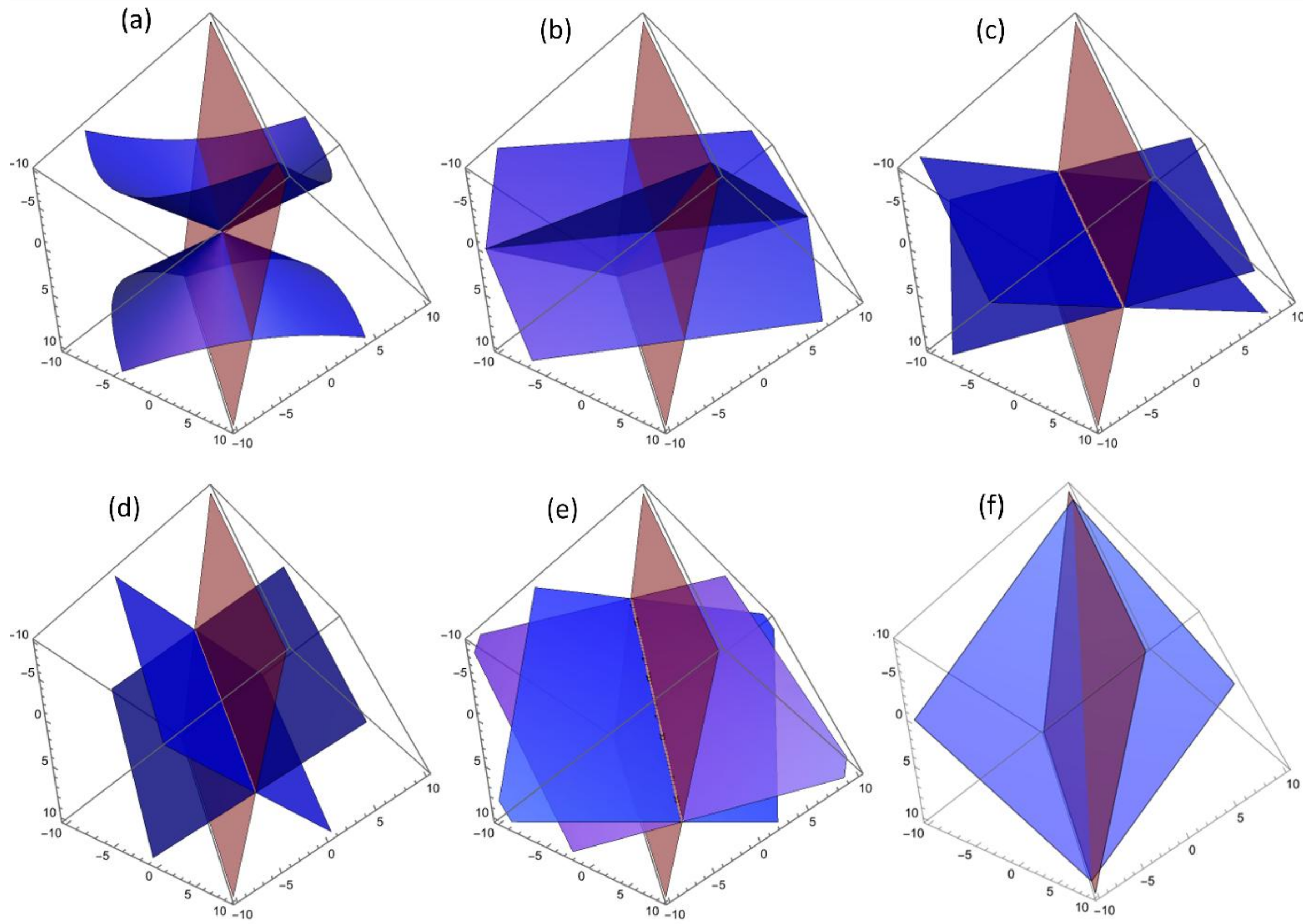

Figure 7. Quadratic CMC surface for the B2-B19' transformation in NiTi alloys with (a) lattice parameters experimentally measured by Kudoh *et al.* [26], $a_{B2} = 3.01$ Å and $a_{B19'} = 2.898$ Å, $b_{B19'} = 4.108$ Å, $c_{B19'} = 4.646$ Å, $\beta = 97.78°$, i.e. $a = 0.9628, b = 1.3648, c = 1.5435$. Oher surfaces are plotted with hypothetical lattice parameters that verify CMC degeneracy condition of first order: (b) $C_1$, with $a = 0.9628, b = \sqrt{2}, c = 1.5435, \beta = 97.78°$, (c) $C_{2a}$, with $a = 1.0166, b = 1.3648, c = 2.1451, \beta = 97.78°$, (d) $C_{2b}$, with $a = 1, b = 1.3142, c = 1.5409, \beta = 90°$, (e) $C_3$, with $a = 0.98, b = 1.4242, c = 1.1767, \beta = 97.78°$. (f) Surface plotted with lattice parameters that verifies degeneracy of second order: $D_1$, with $a = 0.9, b = \sqrt{2}, c = 1.36201, \beta = 97.78°$, which gives the plane $x - y - 2.290\, z = 0$ in blue, and D2 which gives the plane $x + y = 0$ in red. This plane is also the mirror plane of the CMC surfaces whatever the lattice parameters chosen for austenite and martensite (a-f).

The eigenvectors $(\mathbf{eg_1}, \mathbf{eg_2}, \mathbf{eg_3})$ written in columns form the coordinate transformation matrix $\mathbf{P}$. Calculations show that

$$\mathbf{P} = \begin{pmatrix} 1 & 2a^2 - c^2 + \sqrt{\Delta} & 2a^2 - c^2 - \sqrt{\Delta} \\ 1 & -(2a^2 - c^2 + \sqrt{\Delta}) & -(2a^2 - c^2 - \sqrt{\Delta}) \\ 0 & 4\, a\, c\, cos(\beta) & 4\, a\, c\, cos(\beta) \end{pmatrix} \quad (44)$$

with $\Delta = 4\, a^4 + c^4 + 4\, a^2\, c^2 \cos(2\beta)$.

The coordinate transformation is explicitly given by $\begin{bmatrix} x \\ y \\ z \end{bmatrix} = \widetilde{\mathbf{P}} \begin{bmatrix} X \\ Y \\ Z \end{bmatrix}$, where $\widetilde{\mathbf{P}}$ is obtained from $\mathbf{P}$ after normalizing the column vectors. The general expression of $\widetilde{\mathbf{P}}$ is not given here because some signs depend on the values of $a$ and $c$. The eigenvalues $(q_1, q_2, q_3)$ of the CMC matrix are

$$\begin{aligned} q_1 &= \frac{1}{2}(b^2 - 2) \\ q_2 &= \frac{1}{4}(K - \sqrt{\Delta}) \\ q_3 &= \frac{1}{4}(K + \sqrt{\Delta}) \end{aligned} \quad (45)$$

with $K = 2a^2 + c^2 - 4$.

The 6 possible conditions of first order degeneracy deduced from equations (35) are reduced to 3 possible conditions. The details of the calculations are reported in Appendix A. Summarizing the results, the IPS condition (A/M compatibility) is fulfilled for metrics that verify at least one of the following conditions:

$$C_1: \begin{cases} b = \sqrt{2} \\ 2a^2 + c^2 - a^2 c^2 sin^2(\beta) \geq 2 \end{cases}$$

$$C_{2a}: \begin{cases} c = \sqrt{2}\sqrt{\frac{1-a^2}{1-a^2 sin^2(\beta)}} \\ \mathrm{a} \geq 1/sin(\beta) \\ b \leq \sqrt{2} \end{cases}, \quad C_{2b}: \begin{cases} a = 1 \\ \beta = 90° \\ b \leq \sqrt{2} \end{cases}, \text{ or} \quad (46)$$

$$C_3: \begin{cases} c = \sqrt{2}\sqrt{\dfrac{1-a^2}{1-a^2 sin^2(\beta)}} \\ \mathrm{a} \leq 1 \\ b \geq \sqrt{2} \end{cases}$$

Note that the equalities in these conditions can be understood by the fact that the correspondence acts in two distinct subspaces, one along $\mathbf{b}_{B19'}$ and the other one perpendicular to $\mathbf{b}_{B19'}$. Along $\mathbf{b}_{B19'}$, the correspondence $[110]_{B2} \rightarrow [010]_{B19'}$ leads to the condition $b = \sqrt{2}$. The condition $2a^2 + c^2 - a^2c^2sin^2(\beta) = 2$ can be obtained by applying the CMC matrix written in 2D in the plane normal to $\mathbf{b}_{B19'}$, i.e. by considering the planar distortion $(110)_{B2} \rightarrow (010)_{B19'}$.

The degeneracy conditions of second and third orders are given in Appendix B.

For all the NiTi binary alloys reported in literature, the B2 and B19' lattice parameters are such that a $< 1,\ b < \sqrt{2}$, which, according to the equations (46), means that no A/M compatibility can be obtained whatever the value of $c$ or $\beta$. Actually, the fact that a $< 1, b < \sqrt{2}$ results from atomic bonds between Ti atoms created when the B2 structure is transformed into B19', as explained with a hard sphere model [27]. It is concluded that supercompatibility is impossible in binary NiTi alloys, but this not necessarily so in NiTi alloys containing a ternary element. How far however is the actual B19' martensite reported in binary NiTi alloys from a degeneracy condition? More generally, how to determine the distance of the actual B19' in binary NiTi alloys from the IPS conditions that could be targeted in ternary alloys? We propose to introduce two "distances" based on the CMC degeneracy equations (46), one distance for the equality, and the other distance for the inequality. The latter tells how far the metrics is from the frontier obtained when the inequality becomes an equality. If the inequality is not verified, this distance tells how much effort should be made to reach the target, and on the contrary, if the inequality is verified, it permits to estimate the risk to go out of the inequality domain by crossing the frontier when the lattice parameters are changed. The two distances are given in Table 2 for the different IPS conditions for B19' in NiTi alloys.

Table 2. Distances from the equality and inequality conditions required for A/M compatibility (IPS condition) in the case of B2-B19' transformation in binary NiTi alloys with lattice parameters reported by Kudoh *et al.* [26]. The values in green are those that verify the condition, whereas those in red do not.

| Condition | Distance from equality | Inequality to be checked | Distance from inequality frontier |
|---|---|---|---|
| $C_1$ | $\lvert b^2 - 2\rvert$ $= 0.137364$ | $2a^2 + c^2 - a^2c^2sin^2(\beta) \geq 2$ yes | $\lvert 2a^2 + c^2 - a^2c^2sin^2(\beta) - 2\rvert$ $= 0.068402$ |
| $C_{2a}$ | $\left\lvert c^2 - \frac{2(1-a^2)}{1-a^2sin^2(\beta)}\right\rvert$ | $a\ sin(\beta) \geq 1\ and\ b \leq \sqrt{2}$ no | $\lvert a^2sin^2(\beta) - 1\rvert + \lvert b^2 - 2\rvert$ $= 0.227385$ |
| $C_{2b}$ | $\lvert a^2 - 1\rvert + \lvert sin^2(\beta) - 1\rvert$ $= 0.091359$ | $b \geq \sqrt{2}$ no | $\lvert b^2 - 2\rvert$ $= 0.137364$ |
| $C_3$ | $= 0.269706$ | $a \leq 1\ and\ b \geq \sqrt{2}$ no | $\lvert a^2 - 1\rvert + \lvert b^2 - 2\rvert$ $= 0.210399$ |

This table indicates that the most effective method to reach A/M compatibility in a ternary NiTi alloy is to use $C_1$ and try to change the $b$-value such that $b = \frac{b_{B19'}}{a_{B2}}$ reaches $\sqrt{2}$. Since $\lambda_2 = \frac{b_{B19'}}{\sqrt{2}a_{B2}}$ [28], our analysis leads to the same conclusion as that obtained by the PTMC trying to minimize $|\lambda_2 - 1|$. By considering equations (46) with $\beta = 90°$, it can be checked that the same criterion is valid for B19 martensite. As for B19', the conditions $C_{2a}$, $C_{2b}$, and $C_3$ are not reachable for B19 in binary NiTi alloys because its structure and lattice parameters are too close to those of B19' [27]. However, A/M compatible B19 martensite can be formed in NiTi alloys containing other elements. Excellent stability properties such as low thermal and stress-hysteresis, good reversibility and improved fatigue life, have been obtained in various ternary or quaternary NiTi alloys with B19 martensite that nearly verify the IPS condition: NiTiCu [12], NiTiPd [13], NiTiCuPd [29], and NiTiCuCo alloys [30]. Successful results were also reported more recently with B19' martensite on the NiTiCuFe alloys [31].

## 5.2 How far an A/M compatible B19' is from the M/M compatibility?

If the austenite/martensite compatibility (IPS condition) is verified, the CMC equation is degenerated and the habit plane $\mathbf{m}_\mathrm{A}$ can be calculated by equation (37). In addition, the shear direction of an individual variant $\mathbf{d}_\mathbf{A}$ deduced from the shear plane $\mathbf{m}_\mathrm{A}$ by equation (39) is compatible with the twinning system $(\mathbf{a},\mathbf{n})$ if the shear/shear equation (41) is verified. Even if the shear/shear equality (41) is not obtained, it is anyway possible to evaluate how far the metric is from it. Let us give an example in the case with a hypothetical B19' martensite that is A/M compatible. Among the CMC degeneracy conditions of order 1 given in equation (46), let us assume that it is $C_1$ that is fulfilled. We use in this example the same lattice parameters as those in Figure 7b, i.e. $a = 0.9628, b = \sqrt{2}$, $c = 1.5435, \beta = 97.78°$. The CMC quadratic equation (32) $q_{\mathrm{CMC}}(x, y, z) = 0$ is degenerated into a double-plane equation that, after simplification, takes the form:

$$x^2 + y^2 - q_c\, z^2 - 2\, x\, y + q_b\, y\, z + q_b\, x\, z = 0, \text{ with } q_b = 2.10392 \text{ and } q_c = 0.76384$$

This quadratic form can be factorized into the product of two linear equations by noting that the corresponding planes should contain the vector [1,1,0] and the vector $[x, 0,1]$ where $x$ is a solution of the quadratic equation $x^2 + q_b\, x - q_c = 0$. The two possible habit planes are

$$\mathbf{m}_\mathrm{A} = \left( 1, -1, \frac{q_b \pm \sqrt{q_b^2 + 4q_c^2}}{2} \right) \tag{47}$$

Which numerically gives $\mathbf{m}_\mathrm{A}^{+} = (1, -1, 2.41966)$ and $\mathbf{m}_\mathrm{A}^{-} = (1, -1, -0.31568)$. The associated shear directions deduced from the SMC matrix by equation (39) are $\mathbf{d}_\mathrm{A}^{+} = [0.36938, -0.36938, -0.05378]$ and $\mathbf{d}_\mathrm{A}^{-} = [0.115568, -0.115568, 0.216812]$.

The irrational components of the type I and type II transformations twins are recalculated with the equations of section 3.3 with the new lattice parameters. They allow the M/M compatibility. Supercompatibility, i.e. compatibility between the A/M and M/M compatibilities, is obtained if equation (41) is verified. If not (as in the most general cases), the A/M – M/M incompatibility amplitude is evaluated by the ratio

$$\varepsilon = \frac{\left\| 2\, (\mathbf{m}_\mathrm{A}^{\mathrm{t}} \mathbf{n})\, \mathbf{d}_\mathrm{A} - \mathbf{a} \right\|}{s} \tag{48}$$

We recall that $\mathbf{d}_\mathbf{A} = \mathbf{SMC}\, \mathbf{m}_\mathrm{A}$, with $\mathbf{m}_\mathrm{A}$ is the shear plane of the individual martensite variant written in the austenite crystal, $\mathbf{n}$ is the normal to the "shear" twin plane $\mathbf{p}$, and $\mathbf{a}$ is the "shear" twin direction, both vectors written in the austenite crystal. In this formula, the vectors $\mathbf{n}$ and $\mathbf{m}_\mathrm{A}$ are unit vectors in the direct and reciprocal space, respectively. If not, they should be normalized with the austenite metric tensor (see section 3.1). The norm of $\mathbf{a}$ should be the shear amplitude $s$.

Table 3. Transformation twins (shear plane **p**, shear direction **a,** shear amplitude $s$) predicted by the CT for B19’ martensite in NiTi alloys with hypothetical lattice parameters verifying the $C_1$ condition: $a = 0.9628, b = \sqrt{2},\ c = 1.5435, \beta = 97.78°$. The incompatibilities between A/M and M/M compatibilities is evaluated by $\varepsilon$ and $(\widehat{\mathbf{d}_A, \mathbf{a}})$ calculated for the plane $\mathbf{m}_A^+$ or $\mathbf{m}_A^-$ that shows the best compatibilities, i.e. minimum $\varepsilon$. The choice of plane is indicated by the +/- sign in the last column.

| Operator | Two- fold B2 symmetries | Twinning system (p, a) | Twinning shear | Incompatibility amplitude $\varepsilon$ | Incompatibility angle $(\widehat{\mathbf{d}_A, \mathbf{a}})$ | $\mathbf{m}_A^{+/-}$ |
|---|---|---|---|---|---|---|
| $\mathbf{O}_2$ | $m^{B2}_{(001)}, R^{B2}_{\pi,[1\bar{1}0]}$ | Comp.1 : $(100)_{B19'} \parallel (001)_{B2}$<br>$[001]_{B19'} \parallel [\bar{1}10]_{B2}$ | s = 0.27325 | 0.215 | 5.9° | + |
| | $m^{B2}_{(1\bar{1}0)}, R^{B2}_{\pi,[001]}$ | Comp.2 : $(001)_{B19'} \parallel (1\bar{1}0)_{B2}$<br>$[100]_{B19'} \parallel [001]_{B2}$ | | 0.809 | 37.0° | − |
| $\mathbf{O}_4$ | $m^{B2}_{(01\bar{1})}, m^{B2}_{(101)}$ | type I: $(\bar{1}11)_{B19'} \parallel (01\bar{1})_{B2}$<br>$\sim[13, \bar{6}, 19]_{B19'} \parallel [\overline{25}, 13, 13]_{B2}$ | s = 0.30615 | 0.453 | 25.3° | + |
| | $R^{B2}_{\pi,[01\bar{1}]}, R^{B2}_{\pi,[101]}$ | type II: $[\bar{2}11]_{B19'} \parallel [01\bar{1}]_{B2}$<br>$\sim(3\ \bar{4}\ 10)_{B19'} \parallel (\bar{7}\ 3\ 3)_{B2}$ | | 0.566 | 30.0° | − |
| $\mathbf{O}_5$ | $m^{B2}_{(011)}, m^{B2}_{(\bar{1}01)}$ | type I: $(111)_{B19'} \parallel (011)_{B2}$<br>$\sim[\bar{8}\ 5\ 3]_{B19'} \parallel [4, 17, \overline{17}]_{B2}$ | s = 0.17059 | 0.882 | 38.4° | − |
| | $R^{B2}_{\pi,[011]}, R^{B2}_{\pi,[\bar{1}01]}$ | type II: $[211]_{B19'} \parallel [011]_{B2}$<br>$\sim(\bar{8}\ 8\ 7)_{B19'} \parallel (\bar{1}\ 16\ \overline{16})_{B2}$ | | 1.103 | 74.7° | − |
| $\mathbf{O}_6$ | $m^{B2}_{(010)}, m^{B2}_{(100)}$ | type I: $(011)_{B19'} \parallel (010)_{B2}$<br>$\sim[\bar{5}, \bar{2}, 2]_{B19'} \parallel [\bar{7}, 0, \bar{9}]_{B2}$ | s = 0.25502 | 0.469 | 26.9° | − |
| | $R^{B2}_{\pi,[010]}, R^{B2}_{\pi,[100]}$ | type II: $[011]_{B19'} \parallel [010]_{B2}$<br>$\sim(20\ 19\ \overline{19})_{B19'} \parallel (19\ 0\ 20)_{B2}$ | | 0.906 | 64.8° | − |

As shown in Table 3, lattice parameters verifying A/M compatibility (here condition $C_1$) do not generally agree with the twin M/M compatibility, which impedes supercompatibility. It also shows that the transformation twin system closest to supercompatibility is the compound twin noted “Comp.1” in Table 3 for which $\varepsilon = 0.215$.

## 5.3 Ideal lattice parameters of B19’ to get supercompatibility

Is it possible to find lattice parameters of B19’ that could verify both A/M and M/M compatibility conditions? If yes, how many degrees of freedom remain for the B19’ metric? In order to explore this question, a computer program was written in Mathematica. Its input are the set of lattice parameters $a, c, \beta$ that verify the C1 condition ($b = \sqrt{2}$, and $2a^2 + c^2 - a^2c^2sin^2(\beta) \geq 2$). The habit planes $\mathbf{m}_A$ is calculated by equation (47), the shear direction of the transformation $\mathbf{d}_A$ by equation (39), and the incompatibility value $\varepsilon$ by equation (48) with the twin system $(\mathbf{a}, \mathbf{n})$ calculated according to the equations given in section 3.3. The aim of the section is to explain the method without being exhaustive, so only the twin operators $\mathbf{O}_2$ and $\mathbf{O}_4$ were chosen because they already show low value of incompatibility $\varepsilon$ in Table 3. For $\mathbf{O}_2$, the chosen symmetry operations of the twin $\mathbf{p}_A$ and $\mathbf{a}_A$ chosen for the type I and type II twins are $(001)$ and $[001]$, respectively. For $\mathbf{O}_4$, the chosen symmetry operations are $(01\bar{1})$ and $[01\bar{1}]$, respectively. The computer program numerically finds the triplets $a, c, \beta$ that minimize $\varepsilon$. A solution is said supercompatible if $\varepsilon = 0$. The program could find some supercompatible solutions. After analyzing them, we realized that these solutions can be explained, and actually analytically written. We skip the details of this analysis and directly give the conclusions:

For the operator $\mathbf{O}_2$, C1 supercompatible solutions are obtained for the type I twin (actually compound twin) with $\mathbf{p}_\mathrm{A} = (001)$ with the lattice parameters $c = \sqrt{2}$ and $a = \frac{c}{sin(\beta)}$. The habit plane is $\mathbf{m}_\mathrm{A} = (001)$, the shear direction is $\mathbf{d}_\mathrm{A} \parallel [1\bar{1}0]$, the twin shear plane is $\mathbf{p}_\mathrm{A} = (001)$, and the twin shear direction is $\mathbf{a}_\mathrm{A} = 2\,\mathbf{d}_\mathrm{A}$. Supercompatible solutions are also obtained for the type II twin (actually compound twin) with $\mathbf{a}_\mathrm{A} \parallel [001]$, with the lattice parameters $a = 1$ and $c = \frac{\sqrt{2}}{sin(\beta)}$. The habit plane is $\mathbf{m}_\mathrm{A} = (1\bar{1}0)$, the shear direction is $\mathbf{d}_\mathrm{A} \parallel [00\bar{1}]$, the twin shear plane is $\mathbf{jp}_\mathrm{A} = (1\bar{1}0)$ and the twin shear direction is $\mathbf{a}_\mathrm{A} = 2\,\mathbf{d}_\mathrm{A}$. These solutions are actually easy to understand from a geometric point of view; they are illustrated in Figure 8. In these two cases, the habit plane and the junction plane of the twin are parallel; this corresponds to the case described by equation (11), where the martensite transformation is a simple shear, $\delta = 0$, thus $\phi = 0$.

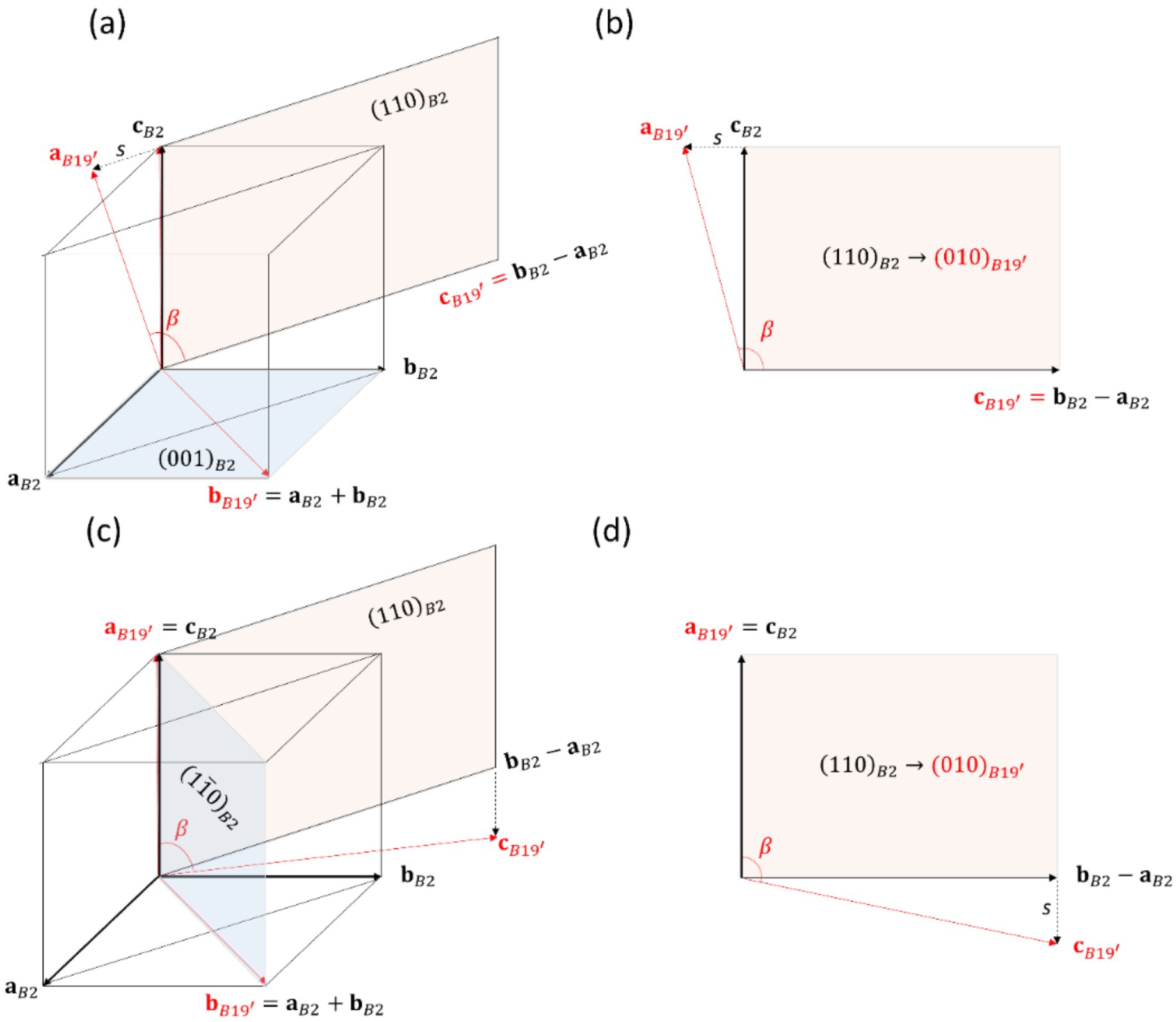

Figure 8. Link between the values of the B2 and B19' lattice parameters in order to get supercompatibility with AM compatibility obtained by C1 condition, and two possible M/M compatibility twins: (a,b) with type I twin by $(001)_{B2}$ reflection, and (c,d) type II twin by $180°/[001]_{B2}$ rotation. (a,c) 3D view of the B2 and B19' lattices. (b,d) $(110)_{B2}$ plane transformed into $(010)_{B19'}$ plane by simple shear. The habit plane $\mathbf{m}_\mathrm{A}$ and the junction plane are identical; it is $(1\bar{1}0)$ in (a) and $(001)$ in (c), in blue.

For the operator $\mathbf{O}_4$, C1 supercompatible solutions for type I twin on the plane $\mathbf{p}_\mathrm{A} = (01\bar{1})$ were obtained numerically with Mathematica. They show that the habit plane is systematically $\mathbf{m}_\mathrm{A} = (1\bar{1}2)$ and the shear direction is $\mathbf{d}_\mathrm{A} \parallel [1\bar{1}\bar{1}]$. By analysing the results, we realized that indeed the B2 plane $\mathbf{m}_\mathrm{A} = (1\bar{1}2)$ can be maintained fully invariant for specific relation between the lattice parameters. The direction $[110]_{B2}$can be maintained invariant when it becomes $\mathbf{b}_{B19'}$ because, as in the case of operator $\mathbf{O}_2$, $b = \sqrt{2}$. In addition, the direction $\mathbf{d}_\mathrm{A} = [\bar{1}11]$ becomes by correspondence $\mathbf{d}_\mathrm{M} = [101]$; it can also be maintained invariant if its norm is unchanged, i.e. if $\sqrt{3} = \sqrt{a^2 + c^2 + 2\,a\,c\cos(\beta)}$. By adding this condition, the supercompatibility conditions were determined analytically. They are

$$a = \frac{1}{2}\sqrt{6 - \frac{4\sqrt{2}}{\tan\beta} - 2\sqrt{1 - \frac{12\sqrt{2}}{\tan\beta}}}$$

$$c = \frac{1}{2}\left( -\cos\beta \sqrt{6 - \frac{4\sqrt{2}\cos\beta}{\sin\beta} - 2\sqrt{1 - \frac{12\sqrt{2}\cos\beta}{\sin\beta}}} + \sqrt{9 + 4\sqrt{2}\cos\beta\sin\beta + \sqrt{1 - \frac{12\sqrt{2}\cos\beta}{\sin\beta}} + \cos(2\beta)\left(3 - \sqrt{1 - \frac{12\sqrt{2}\cos\beta}{\sin\beta}}\right)} \right)$$

The values of $a$ and $c$ are functions of $\beta$; they are shown in Figure 9. The supercompatible subspace of the metric space is thus 1-dimensional. The supercompatible values obtained for a monoclinic angle close to that of the actual B19' in binary NiTi alloys, $\beta = 98°$, are $a = 0.8825$, $c = 1.6182$, which gives $a_{B19'} = 2.65$ Å and $c_{B19'} = 4.85$ Å for $a_{B2} = 3.0$ Å. Such values are not far aways ($< 8\%$) from the actual values of B19' in NiTi; they can thus be probably targeted by adding ternary or quaternary elements to the alloy.

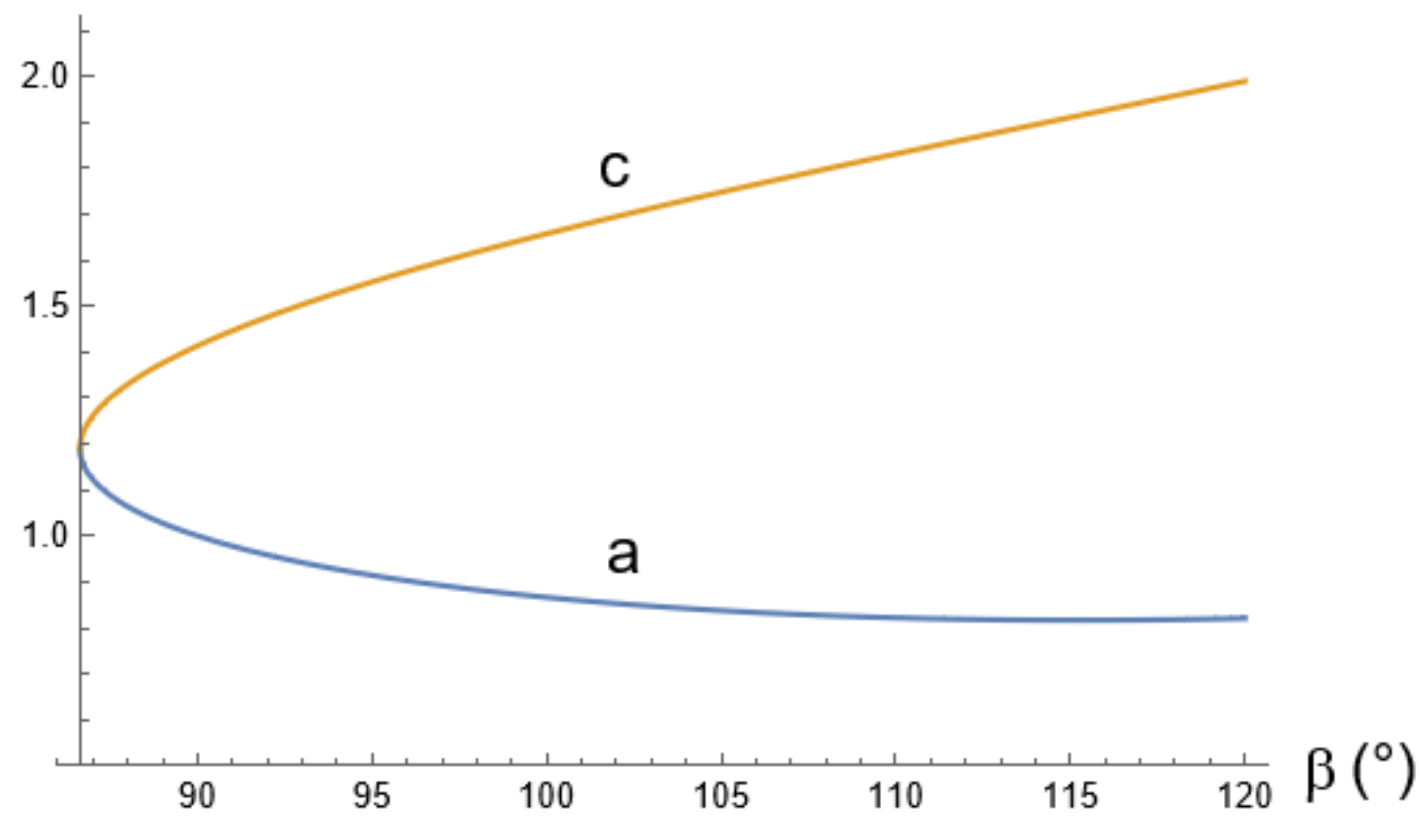

Figure 9. Curves of $a$ and $c$ as functions of $\beta$ in order to get supercompatibility with C1 A/M compatibility and M/M compatibility by type I twin on $(01\bar{1})_{B2}$ reflection. The two curves join, $a = c$, at the angle $\beta = 86.6°$.

# 6 Discussion

In this paper, we have shown that supercompatibility is obtained between the austenite matrix and a laminate constituted of two martensite variants when the A/M interface is coherent, i.e. the $A \rightarrow M$ lattice distortion is an IPS, when the M/M interface is also coherent, i.e. the two martensite variants are linked by type I or type II twin, and when shear direction of the IPS is proportional to the shear direction of the transformation twin. These conditions can be mathematically expressed directly from the correspondence matrix, the metric tensors and symmetry groups of austenite and martensite. Indeed, the transformation twins (type I and II) are deduced from the symmetry elements of the austenite by double-coset decomposition based on the correspondence intersection subgroup, as detailed in section 3.3. The shear plane of normal $\mathbf{n}$, shear amplitude $s$, and shear direction $\mathbf{a}$ of type I twins are given by equations (17), (18), and (19), respectively. The shear direction $\mathbf{a}$, shear amplitude $s^*$, and shear plane of normal $\mathbf{n}$ of type II twins are given by equations (21),(22), and (23), respectively. The A/M compatibility (IPS condition) is obtained when the quadratic equation of the CMC double-cone is degenerated into that of a double plane. The degeneracy condition is given by equation (35), and the Miller indices of the habit plane $\mathbf{m}_A$ by equation (37). The A/M/M supercompatibility is then obtained when the shear direction $\mathbf{d}_A$ deduced from $\mathbf{m}_A$ by the SMC matrix with equation (39) is proportional to the shear direction $\mathbf{a}$ of the transformation twins by equation (41). The main compatibility equations are thus the twin equations (17)-(23) for the M/M compatibility, the degeneracy equation (35) for the A/M compatibility, and the shear/shear equation to make these two types of compatibilities compatible (41). If we compare them with the SC equations (1)-(3), some similarities can be recognized, but the expected formal equivalence is not obvious. Demonstrating the equivalence is beyond the aim of the paper; only some elements of response are proposed. The main one is the equation that links $\mathbf{U}$ to the correspondence matrix and metric tensors. Any direction of the austenite crystal $\mathbf{u}_{\mathrm{A}}$ becomes by distortion a vector $\mathbf{u'}_{\mathrm{A}} = \mathbf{F}\,\mathbf{u}_{\mathrm{A}}$, with $\mathbf{F} = \mathbf{R}\,\mathbf{U}$, and this image vector is a direction of martensite $\mathbf{u}_{\mathrm{M}} = \mathbf{C}^{\mathrm{M}\to\mathrm{A}}\,\mathbf{u}_{\mathrm{A}}$. The square of the norm of $\mathbf{u'}_{\mathrm{A}}$ is $\|\,\mathbf{u'}_{\mathrm{A}}\|^2 = \mathbf{u'}^{\mathrm{t}}_{\mathrm{A}}\,\boldsymbol{\mathcal{M}}_{\mathrm{A}}\,\mathbf{u'}_{\mathrm{A}} = \mathbf{u}^{\mathrm{t}}_{\mathrm{A}}\,\mathbf{U}^{\mathrm{t}}\,\mathbf{R}^{\mathrm{t}}\,\boldsymbol{\mathcal{M}}_{\mathrm{A}}\mathbf{R}\,\mathbf{U}\,\mathbf{u}_{\mathrm{A}} = \mathbf{u}^{\mathrm{t}}_{\mathrm{A}}\,\mathbf{U}^{\mathrm{t}}\,\boldsymbol{\mathcal{M}}_{\mathrm{A}}\,\mathbf{U}\,\mathbf{u}_{\mathrm{A}}$ because $\mathbf{R}^{\mathrm{t}}\,\boldsymbol{\mathcal{M}}_{\mathrm{A}}\mathbf{R} = \boldsymbol{\mathcal{M}}_{\mathrm{A}}$ by equation (16). Since $\|\,\mathbf{u'}_{\mathrm{A}}\|^2 = \|\,\mathbf{u}_{\mathrm{M}}\|^2 = \mathbf{u}^{\mathrm{t}}_{\mathrm{M}}\,\boldsymbol{\mathcal{M}}_{\mathrm{M}}\,\mathbf{u}_{\mathrm{M}} = \mathbf{u}^{\mathrm{t}}_{\mathrm{A}}\left(\mathbf{C}^{\mathrm{M}\to\mathrm{A}}\right)^{\mathrm{t}}\,\boldsymbol{\mathcal{M}}_{\mathrm{M}}\,\mathbf{C}^{\mathrm{M}\to\mathrm{A}}\,\mathbf{u}_{\mathrm{A}}$, we get

$$\mathbf{U}^{\mathrm{t}}\,\boldsymbol{\mathcal{M}}_{\mathrm{A}}\,\mathbf{U} = \left(\mathbf{C}^{\mathrm{M}\to\mathrm{A}}\right)^{\mathrm{t}}\,\boldsymbol{\mathcal{M}}_{\mathrm{M}}\,\mathbf{C}^{\mathrm{M}\to\mathrm{A}} \qquad (49)$$

If the austenite phase is cubic with a lattice parameter $a_A$, $\boldsymbol{\mathcal{M}}_{\mathrm{A}} = a_A^2\,\mathbf{I}$, and equation (49) becomes $a_A^2\,\mathbf{U}^2 = \left(\mathbf{C}^{\mathrm{M}\to\mathrm{A}}\right)^{\mathrm{t}}\,\boldsymbol{\mathcal{M}}_{\mathrm{M}}\,\mathbf{C}^{\mathrm{M}\to\mathrm{A}}$; consequently $\mathbf{CMC} = a_A^2\,(\mathbf{U}^2 - \mathbf{I})$ and $\mathbf{SMC} = \frac{1}{a_A^2}\,(\mathbf{I} - \mathbf{U}^{-2})$. The matrix $(\mathbf{U}^2 - \mathbf{I})$ appears in the SC equation (2), but no clear equivalence between the CT and SC equations could be yet formally demonstrated. For the moment, we could just check numerically that in all the cases of supercompatible lattice parameters of B19' determined from the CT equations with (001) type I, [001] type II, or (01-1) type I twins in section 5.3, the stretch matrix $\mathbf{U}$ deduced from $\mathbf{U}^2$ calculated by equation (49) verifies exactly the SC equations (1)-(3). Therefore, even if formal proof needs to be given, we think that the CT and PTMC approaches are equivalent. The advantages of the CT is however that it does not require calculating the stretch matrices; its input are the elementary crystallography inputs: symmetries and parameters of the austenite and martensite lattices, and correspondence between them. The CT approach is based on simple geometry, which makes the M/M twin equations, the A/M compatibility equation, and the shear/shear equation probably easier to understand than the SC equations.

# 7 Conclusions

The PTMC gives the equations to calculate the transformation twins between the martensite variants, the habit planes between austenite and bi-variant laminates, and more recently, the supercompatibility conditions. These conditions are written as a set of three (SC) equations, all based on the stretch matrix $\mathbf{U}$. The first one, $\lambda_2 = 1$, means that a free rotation can be combined with the stretch matrix to make the A → M lattice distortion an IPS. The second and third ones, called cofactor conditions, depend on the twinning mode that links the variants together. They are currently understood as additional conditions that allow for the formation of bi-variant laminate martensite products with a coherent interface with austenite whatever the volume fraction of each variant in the products. The SC conditions are currently used to design new shape memory alloys with improved cyclability and fatigue resistance. In this paper, we proposed a different approach, called correspondence theory (CT) that gives the same results as the PTMC, but it is based elementary crystallography tools: symmetry groups, lattice parameters, and correspondence. Note that the CT can treat the cases of non-cubic austenite, which seems more difficult with the PTMC because polar decomposition and stretch matrices require the introduction of an orthonormal basis.

First, it was shown with simple geometric arguments that if the A/M interface is coherent, i.e. the A → M lattice distortion is an IPS, and if two variants $M_1$ and $M_2$ are linked by a transformation twin, then there is a way to position the A/M interfaces of $M_1$ and $M_2$ such that the IPS shear direction of $M_1$ and that of $M_2$ come in coincidence and allows for a perfect coherency at the $M_1$/$M_2$ twin plane. For such cases, the common IPS shear direction and the twin shear direction are proportional and linked by a "shear/shear" equation. This equation makes the A/M compatibility and the M/M compatibility compatible, and the A/M/M system becomes supercompatible.

The CT permits to calculate the transformation twins from the correspondence and symmetries. First, the correspondence intersection subgroup is first determined, and used to partition the austenite symmetry group into a) left cosets that define the different correspondence variants, and b) double-cosets that define the different intercorrespondences between the variants. Each intercorrespondence is made of different austenite symmetry elements. Among them, the reflections and the 180° rotation symmetries become by correspondence transformation twins of type I and type II, respectively. The mirror plane of the type I twin (shear plane) and the 180° rotation axis of the type II twin (shear direction) are generic; they do not depend on the metrics. The shear direction of the type I twin and the shear plane of the type II twin are not generic, such as the shear amplitude. The CT gives the formulae to calculate them directly from the correspondence matrix, symmetry matrix, and metrics.

The CT permits to calculate the lattice parameters required to obtain A/M compatibility, i.e. an A → M IPS lattice distortion. A symmetric matrix called $\mathbf{CMC}$ for "compatibility of metrics by correspondence" was introduced. The A/M compatibility is obtained when the CMC quadratic equation of double-cone is degenerated into that of a double-plane (first-order degeneracy), a plane (second-order degeneracy), or the full space (third-order degeneracy). The A/M compatibility condition can thus be written as simple conditions on the eigenvalues $q_i$ of the $\mathbf{CMC}$ matrix: $q_i = 0$ & $q_j\, q_k \leq 0$, and the A/M habit plane is calculated from $q_j$ and $q_k$. The CT permits to determine the shear direction of the A → M lattice distortion from the A/M habit plane. The topological subspace of the lattice parameters allowing A/M compatibility can then be further reduced to that of supercompatibility by considering the shear/shear equation.

Different distances between actual and ideal martensite lattice parameters that could be targeted for A/M compatibilities and A/M/M supercompatibility can be calculated for different transformation twinning systems. Examples are given in the case of B19' martensite in NiTi alloys. It was checked numerically that the supercompatible solutions obtained by the CT also verify the SC equations. However, a formal equivalence between the PTMC and the CT could not yet be stablished.

The CT gives a geometric understanding of the crystallography of martensitic transformations; it makes quite straightforward the calculations of the transformation twins and supercompatibility conditions, and can thus be useful for phase engineering new shape memory alloys.

## Acknowledgments

I would like to acknowledge Prof. Tomonari Inamura for the discussions on the manuscript. Some errors of interpretation in the initial version have been corrected thanks to his careful reading.

## Comments

The author did not use artificial intelligence for neither bibliography, discussion or writing. The algebraic expressions were determined by hand and with Mathematica. Mathematica was also used for the numerical calculations and for plotting the 2D lattices and 3D surfaces. The text was written with Word LTSC, with references inserted by Zotero.

# Appendix A: Calculations of the A/M compatibility conditions for B19' martensite in NiTi alloys by degeneracy of first order

The 6 possible conditions of first order degeneracy of the CMC quadratic form deduced from equations (35) are reduced to 3 possible conditions $C_1$, $C_2$ or $C_3$, that are

---

$$C_1:\ \begin{cases} q_1 = 0 \\ q_2 q_3 \le 0 \end{cases} \Leftrightarrow \begin{cases} b = \sqrt{2} \\ K^2 - \Delta \le 0 \end{cases} \Leftrightarrow \begin{cases} b = \sqrt{2} \\ 2a^2 + c^2 - a^2 c^2 sin^2(\beta) \ge 2 \end{cases}$$

---

$$C_2:\ \begin{cases} q_2 = 0 \\ q_1 q_3 \le 0 \end{cases} \Leftrightarrow \begin{cases} \mathrm{K} = \sqrt{\Delta} \\ (b^2 - 2)(K + \sqrt{\Delta}) \le 0 \end{cases} \Leftrightarrow \begin{cases} \mathrm{K} = \sqrt{\Delta} \\ b^2 \le 2 \end{cases}$$

$$\Leftrightarrow \begin{cases} K^2 - \Delta = 0 \\ K \ge 0 \\ b \le \sqrt{2} \end{cases} \Leftrightarrow \begin{cases} 2a^2 + c^2 - a^2 c^2 sin^2(\beta) = 2 \\ 2a^2 + c^2 \ge 4 \\ b \le \sqrt{2} \end{cases} \Leftrightarrow \begin{cases} 2a^2 + c^2 - a^2 c^2 sin^2(\beta) = 2 \\ a^2 c^2 sin^2(\beta) \ge 2 \\ b \le \sqrt{2} \end{cases}$$

If $a \neq 1/sin(\beta)$, the first equality gives a solution only for $a \le 1$ or $a > 1/sin(\beta)$ that is $c = \sqrt{2}\sqrt{\frac{1-a^2}{1-a^2 sin^2(\beta)}}$. The case $a = 1/sin(\beta)$ leads to $a = 1$, and is thus possible only if $\beta = 90°$. The second equation (inequality) can be written by substituting $c$ as two conditions: (a) or (b) with

$$\left| \begin{array}{l} (a):\ sin^2(\beta)\, a^4 - 2sin^2(\beta) a^2 + 1 \le 0 \text{ if } a < 1/sin(\beta) \\ (b):\ sin^2(\beta)\, a^4 - 2sin^2(\beta) a^2 + 1 \ge 0 \text{ if } a > 1/sin(\beta) \end{array} \right.$$

The condition (a) is verified only if $a = 1\ \&\ \beta = 90°$. The condition (b) is always verified. Consequently:

$$C_2 \Leftrightarrow \begin{cases} c = \sqrt{2}\sqrt{\dfrac{1-a^2}{1-a^2 sin^2(\beta)}} \\ a \ge 1/sin(\beta) \\ b \le \sqrt{2} \end{cases} or \begin{cases} a = 1 \\ \beta = 90° \\ b \le \sqrt{2} \end{cases}$$

---

$$C_3:\ \begin{cases} q_3 = 0 \\ q_1 q_2 \le 0 \end{cases} \Leftrightarrow \begin{cases} \mathrm{K} = -\sqrt{\Delta} \\ (b^2 - 2)(K - \sqrt{\Delta}) \le 0 \end{cases} \Leftrightarrow \begin{cases} \mathrm{K} = -\sqrt{\Delta} \\ (b^2 - 2) \ge 0 \end{cases}$$

$$\Leftrightarrow \begin{cases} K^2 - \Delta = 0 \\ K \le 0 \\ b \ge \sqrt{2} \end{cases} \Leftrightarrow \begin{cases} 2a^2 + c^2 - a^2 c^2 sin^2(\beta) = 2 \\ 2a^2 + c^2 \le 4 \\ b \ge \sqrt{2} \end{cases} \Leftrightarrow \begin{cases} 2a^2 + c^2 - a^2 c^2 sin^2(\beta) = 2 \\ a^2 c^2 sin^2(\beta) \le 2 \\ b \ge \sqrt{2} \end{cases}$$

By using the same arguments as for $C_2$, we obtain

$$C_3 \Leftrightarrow \begin{cases} c = \sqrt{2}\sqrt{\dfrac{1-a^2}{1-a^2 sin^2(\beta)}} \\ a \le 1 \\ b \ge \sqrt{2} \end{cases}$$

---

# Appendix B: A/M compatibility conditions for B19' martensite in NiTi alloys by degeneracy of second and third order

The conditions of degeneracy of second order are found by solving equations (36). The conditions $D_1$ given by $q_1 = q_2 = 0$ or $q_1 = q_3 = 0$ leads to $b = \sqrt{2}$ and $K^2 = \Delta$ . The conditions $D_2$ given by $q_2 = q_3 = 0$ leads to $K^2 = \Delta = 0$, which implies that $2a^2 + c^2 = 4$ and $a^2c^2sin^2(\beta) = 2$, which is possible only if $\beta = 90°$ and $a = 1$. Thus, the conditions of second order degeneracy are

$$D_1 : \begin{cases} c = \sqrt{2}\sqrt{\dfrac{1-a^2}{1-a^2 sin^2(\beta)}} \\ a < 1 \; OR \; \mathrm{a} \geq \dfrac{1}{sin(\beta)} \\ b = \sqrt{2} \end{cases} \quad \text{or} \quad D_2 : \begin{cases} a = 1 \\ c = \sqrt{2} \\ \beta = 90° \end{cases} \tag{50}$$

The degeneracy condition $D_1$ gives *a priori* two planes in the eigen basis: $Y = 0$ or $Z = 0$. Calculations with the coordinate transformation matrix **P** given by equation (44) show that any vector $[X, 0, Z]$ in the plane $Y = 0$ is written in the austenite basis $(x, y, z)$ that are such that $x - y - 2\sqrt{\frac{1-a^2}{2-c^2}}z = 0$ for $a < 1$ or $x - y + 2\sqrt{\frac{1-a^2}{2-c^2}}z = 0$ for $\mathrm{a} \geq \frac{1}{sin(\beta)}$. Calculations with $Z = 0$ lead to the same planes. Consequently, for degeneracy of second order $D_1$, the double-plane is reduced to the simple plane $(1, -1, \mp 2\sqrt{\frac{1-a^2}{2-c^2}})$ and the sign of the last Miller index depends on the value of $a$.

The degeneracy plane for the condition $D_2$ is $X = 0$. It is the plane normal to the eigenvector $\mathbf{eg_1}$ . Its Miller indices in the crystallographic basis are (1,1,0); it is the plane $x + y = 0$.

The condition of degeneracy of third order can be found by solving equation (38), or by double derivative, $\frac{\partial^2 q_{\mathrm{CMC}}(x,y,z)}{\partial x^2} = \frac{\partial^2 q_{\mathrm{CMC}}(x,y,z)}{\partial y^2} = 0 \Rightarrow b^2 + c^2 = 4$, $\frac{\partial^2 q_{\mathrm{CMC}}(x,y,z)}{\partial z^2} = 0 \Rightarrow a = 1$, $\frac{\partial^2 q_{\mathrm{CMC}}(x,y,z)}{\partial xy} = 0 \Rightarrow b = c$, and $\frac{\partial^2 q_{\mathrm{CMC}}(x,y,z)}{\partial xz} = 0 \Rightarrow \cos(\beta) = 0$. These equations are thus reduced to

$$E : \begin{cases} a = 1 \\ b = c = \sqrt{2} \\ \beta = 90° \end{cases} \tag{51}$$

The condition *E* corresponds to the case where the metrics of austenite and martensite are in perfect correspondence, i.e. the matrix CMC is null, and all the vectors $(x, y, z)$ of $\mathbb{R}^3$, not just a unique plane, verify equation (30).

# References

[1] J.S. Bowles, J.K. Mackenzie, The crystallography of martensite transformations I, Acta Metallurgica 2 (1954) 129–137. https://doi.org/10.1016/0001-6160(54)90102-9.

[2] M.S. Wechsler, D.S. Lieberman, T.A. Read, On the Theory of the Formation of. Martensite, Transactions AIME 197 (1953) 1503–1515.

[3] C.M. Wayman, The Growth of Martensite Since E.C. Bain (1924) - Some Milestones, MSF 56–58 (1990) 1–32. https://doi.org/10.4028/www.scientific.net/MSF.56-58.1.

[4] K. Bhattacharya, Microstructure of martensite. Why it forms and how it gives rise to the shape-memory effect., Oxford University Press, 2012.

[5] E.C. Bain, The nature of martensite, Trans. AIME 70 (1924) 25–46.

[6] H.K.D.H. Bhadeshia, A tribute to Professor “Jack” Christian, J. Phys. IV France 112 (2003) 17–25. https://doi.org/10.1051/jp4:2003835.

[7] J.W. Christian, The Theory of Transformations in Metals and Alloys (2nd ed.), Pergamon Press (Elsevier), 1975.

[8] J.M. Ball, R.D. James, Fine phase mixtures as minimizers of energy, Archive for Rational Mechanics and Analysis 100 (1987) 13–52.

[9] J.M. Ball, Mathematical models of martensitic microstructure, Materials Science and Engineering: A 378 (2004) 61–69. https://doi.org/10.1016/j.msea.2003.11.055.

[10] K. Otsuka, X. Ren, Physical metallurgy of Ti–Ni-based shape memory alloys, Progress in Materials Science 50 (2005) 511–678. https://doi.org/10.1016/j.pmatsci.2004.10.001.

[11] P. Chowdhury, H. Sehitoglu, Deformation physics of shape memory alloys – Fundamentals at atomistic frontier, Progress in Materials Science 88 (2017) 49–88. https://doi.org/10.1016/j.pmatsci.2017.03.003.

[12] J. Cui, Y.S. Chu, O.O. Famodu, Y. Furuya, Jae. Hattrick-Simpers, R.D. James, A. Ludwig, S. Thienhaus, M. Wuttig, Z. Zhang, I. Takeuchi, Combinatorial search of thermoelastic shape-memory alloys with extremely small hysteresis width, Nature Mater 5 (2006) 286–290. https://doi.org/10.1038/nmat1593.

[13] R. Delville, D. Schryvers, Z. Zhang, R. James, Transmission electron microscopy investigation of microstructures in low-hysteresis alloys with special lattice parameters, Scripta Materialia 60 (2009) 293–296. https://doi.org/10.1016/j.scriptamat.2008.10.025.

[14] R.D. James, Z. Zhang, A Way to Search for Multiferroic Materials with “Unlikely” Combinations of Physical Properties, in: Magnetism and Structure in Functional Materials, Springer, 2004: pp. 159–174.

[15] H. Gu, L. Bumke, C. Chluba, E. Quandt, R.D. James, Phase engineering and supercompatibility of shape memory alloys, Materials Today 21 (2018) 265–277. https://doi.org/10.1016/j.mattod.2017.10.002.

[16] Z. Zhang, R.D. James, S. Müller, Energy barriers and hysteresis in martensitic phase transformations, Acta Materialia 57 (2009) 4332–4352. https://doi.org/10.1016/j.actamat.2009.05.034.

[17] X. Chen, V. Srivastava, V. Dabade, R.D. James, Study of the cofactor conditions: Conditions of supercompatibility between phases, Journal of the Mechanics and Physics of Solids 61 (2013) 2566–2587. https://doi.org/10.1016/j.jmps.2013.08.004.

[18] Y. Song, X. Chen, V. Dabade, T.W. Shield, R.D. James, Enhanced reversibility and unusual microstructure of a phase-transforming material, Nature 502 (2013) 85–88. https://doi.org/10.1038/nature12532.

[19] C. Cayron, The Correspondence Theory and Its Application to NiTi Shape Memory Alloys, Crystals 12 (2022) 130. https://doi.org/10.3390/cryst12020130.

[20] C. Cayron, The transformation matrices (distortion, orientation, correspondence), their continuous forms and their variants, Acta Crystallogr A Found Adv 75 (2019) 411–437. https://doi.org/10.1107/S205327331900038X.

[21] C. Cayron, The concept of axial weak twins, Acta Materialia 236 (2022) 118128. https://doi.org/10.1016/j.actamat.2022.118128.
[22] M. Bevis, Crocker A.G., Twinning modes in lattices, Proc. Roy. Soc. Lond. A 313 (1969) 509–529.
[23] C. Cayron, Groupoid of orientational variants, Acta Crystallogr A Found Crystallogr 62 (2006) 21–40. https://doi.org/10.1107/S010876730503686X.
[24] V. Janovec, Group analysis of domains and domain pairs, Czech J Phys 22 (1972) 974–994. https://doi.org/10.1007/BF01690203.
[25] V. Janovec, Th. Hahn, H. Klapper, Twinning and domain structures, in: 2013: pp. 397–412. https://doi.org/10.1107/97809553602060000916.
[26] Y. Kudoh, M. Tokonami, S. Miyazaki, K. Otsuka, Crystal structure of the martensite in Ti-49.2 at.%Ni alloy analysed by the single crystal X-ray diffraction method, (n.d.).
[27] C. Cayron, Hard-sphere model of the B2 → B19' phase transformation, and its application to predict the B19' structure in NiTi alloys and the B19 structures in other binary alloys, Acta Materialia 270 (2024) 119870. https://doi.org/10.1016/j.actamat.2024.119870.
[28] R.D. James, K.F. Hane, Martensitic transformations and shape memory materials, Acta Mater. 48 (2000) 197-222.
[29] X.L. Meng, H. Li, W. Cai, S.J. Hao, L.S. Cui, Thermal cycling stability mechanism of Ti50.5Ni33.5Cu11.5Pd4.5 shape memory alloy with near-zero hysteresis, Scripta Materialia 103 (2015) 30–33. https://doi.org/10.1016/j.scriptamat.2015.02.030.
[30] Z. Yang, D. Cong, Y. Yuan, R. Li, H. Zheng, X. Sun, Z. Nie, Y. Ren, Y. Wang, Large room-temperature elastocaloric effect in a bulk polycrystalline Ni-Ti-Cu-Co alloy with low isothermal stress hysteresis, Applied Materials Today 21 (2020) 100844. https://doi.org/10.1016/j.apmt.2020.100844.
[31] H. Zhang, J. Liu, Z. Ma, Y. Ren, D. Jiang, L. Cui, K. Yu, Small stress-hysteresis in a nanocrystalline TiNiCuFe alloy for elastocaloric applications over wide temperature window, Journal of Alloys and Compounds 928 (2022) 167195. https://doi.org/10.1016/j.jallcom.2022.167195.